\documentclass[12pt]{article}
\pdfoutput=1
\usepackage{putex}
\usepackage{comment}
\usepackage{graphicx}
\usepackage{caption}
\usepackage{amsmath}
\usepackage{amssymb}
\usepackage[usenames,dvipsnames,table]{xcolor}
\usepackage{array}
\usepackage{subcaption}
\usepackage{epstopdf}
\usepackage{enumerate}
\usepackage{cite}
\usepackage{tensor}
\usepackage{slashed}
\usepackage[export]{adjustbox}
\usepackage[aligntableaux=center]{ytableau}
\usepackage{dsfont}
\usepackage[utf8]{inputenc}
\usepackage[
      colorlinks=true,
      linkcolor=blue,
      urlcolor=blue,
      filecolor=black,
      citecolor=red,
      ]{hyperref}
\newcommand{\RNum}[1]{\uppercase\expandafter{\romannumeral #1\relax}}

\newcommand {\be} {\begin {equation}}
\newcommand {\ee} {\end {equation}}

\newcommand {\bes} {\begin {equation*}}
\newcommand {\ees} {\end {equation*}}

\newcommand{\cD}{{\mathcal D}}

\newcommand{\beq}{\begin{equation}}
\newcommand{\eeq}{\end{equation}}

\def\ie{\begin{equation}\begin{aligned}}
\def\fe{\end{aligned}\end{equation}}

\numberwithin{equation}{section}

\def\<{\langle}
\def\>{\rangle}

\usepackage{tikz-feynman}

\def\halfd{\tfrac{d}{2}}

\def\co{\mathcal{O}}
\def\cD{\mathcal{D}}

\newcommand\braket[1]{\langle #1 \rangle}

\newcommand\lr[1]{\left(#1\right)}
\def\Seff{\frac{1}{2}\log\det[\frac{C}{|y-z|^{d+s}}+\sigma(y) \delta^d(y-z)]-\hat{j}\log(G(x_1,x_2;\sigma))}

\def\ba#1\ea{%
    \begin{align}%
    #1%
    \end{align}%
}  

\def\bfla#1\efla{%
    \begin{flalign*}%
    #1%
    \end{flalign*}%
}  

\def\beq#1\eeq{%
    \begin{equation}%
    #1%
    \end{equation}%
}

\begin{document}

\preprint{PUPT-2633}

\institution{PU}{Department of Physics, Princeton University, Princeton, NJ 08544, USA}

\title{Long Range, Large Charge, Large $N$} 

\authors{Simone Giombi, Elizabeth Helfenberger, and Himanshu Khanchandani}

\abstract{We study operators with large charge $j$ in the $d$-dimensional $O(N)$ model with long range interactions that decrease with the distance as $1/r^{d+s}$, where $s$ is a continuous parameter. We consider the double scaling limit of large $N$, large $j$ with $j/N=\hat{j}$ fixed, and identify the semiclassical saddle point that captures the two-point function of the large charge operators in this limit. The solution is given in terms of certain ladder conformal integrals that have recently appeared in the literature on fishnet models. We find that the scaling dimensions for general $s$ interpolate between $\Delta_j \sim \frac{(d-s)}{2}j$ at small $\hat{j}$ and $\Delta_j \sim \frac{(d+s)}{2}j$ at large $\hat{j}$, which is a qualitatively different behavior from the one found in the short range version of the $O(N)$ model. We also derive results for the structure constants and 4-point functions with two large charge and one or two finite charge operators. Using a description of the long range models as defects in a higher dimensional local free field theory, we also obtain the scaling dimensions in a complementary way, by mapping the problem to a cylinder in the presence of a chemical potential for the conserved charge.}

\date{}
\maketitle

\tableofcontents

\section{Introduction and Summary}
Long range $O(N)$ models are interesting generalizations of the familiar ``short range" $O(N)$ symmetric spin systems. While in the latter the spins only have nearest-neighbor interactions, in the long range models all spins interact with each other with a strength that depends on the distance $r$ as a power law $\sim \frac{1}{r^{\alpha}}$. The exponent $\alpha$ is usually parameterized as $\alpha=d+s$, where $d$ is the spacetime dimension and $s$ a real parameter. The long range models have second order phase transitions over a range of $s$, with critical exponents being non-trivial functions of this continuous parameter.  
Vector models with long range interactions have a long history \cite{PhysRevLett.29.917, PhysRevB.8.281, PhysRevB.15.4344}, and various aspects of their physics have also been revisited in several recent works \cite{Paulos:2015jfa, Gubser:2017vgc, Behan:2017dwr, Behan:2017emf, Behan:2018hfx, Gubser:2019uyf, Giombi:2019enr, Chai:2021arp, Chakraborty:2021lwl, Chai:2021tpt}. In this paper, we focus on the spectrum of operators that carry a large charge under the $O(N)$ symmetry. CFT dynamics simplify significantly when considering operators with a large charge under some global symmetry, as has been observed extensively in the last few years \cite{Hellerman:2015nra, Alvarez-Gaume:2016vff, Monin:2016jmo, Badel:2019oxl, Alvarez-Gaume:2019biu, Badel:2019khk, Cuomo:2021ygt, Cuomo:2020rgt, Antipin:2020abu, Cuomo:2021cnb, Moser:2021bes, Orlando:2021usz, Dondi:2022wli} (also see \cite{Gaume:2020bmp} for a review and more references). 

In the continuum limit, the long range $O(N)$ model may be described by an action containing a non-local kinetic term and a local quartic interaction term \cite{PhysRevLett.29.917, Gubser:2017vgc, Giombi:2019enr}
\begin{equation} \label{ActionLongRangeQuadratic}
S= \frac{C}{2}\int d^d x d^d y \dfrac{\phi^I(x) \phi^I (y)}{|x-y|^{d+s}} + \dfrac{g}{4} \int d^d x (\phi^I \phi^I (x))^2, \qquad C = \dfrac{2^{s} \Gamma(\frac{d+s}{2})}{\pi^{d/2}\Gamma(-\frac{s}{2})},
\end{equation}
where we work in the Euclidean flat space $R^d$. 
The scaling dimension of the fundamental field is $\Delta_{\phi}=\frac{d-s}{2}$ and does not get renormalized due to the non-local nature of the quadratic term (composite operators, on the other hand, can have non-trivial anomalous dimensions). It is well-known that the model has nontrivial RG fixed points in the range $d/2 < s < s^*$. For $s< d/2$, where the quartic term becomes irrelevant, the low energy behavior of the model is described by the Gaussian (generalized free field) fixed point, while above the upper critical value $s_*$  it is described by the usual short range $O(N)$ symmetric fixed point. The critical value is $s^*=2-2\gamma_{\phi}^{\rm SR}$, where $\gamma_{\phi}^{\rm SR}$ is the anomalous dimension of the fundamental field at the short range fixed point (the value of $s^*$ is such that the scaling dimension $\Delta_{\phi}$ at the long range fixed point becomes equal to that of the short range fixed point). 
 Near the lower limit of the range of $s$, i.e. for $s = (d + \epsilon)/2$, the model has a perturbative Wilson-Fisher fixed point with $g \sim O(\epsilon)$ \cite{PhysRevLett.29.917}, while a weakly coupled description near the upper limit $s^*$ was recently proposed in \cite{Behan:2017dwr, Behan:2017emf}. 

In this paper we will focus on the large $N$ limit of the long range theory (\ref{ActionLongRangeQuadratic}). In this limit, it is convenient to introduce a Hubbard-Stratonovich auxiliary field in a way analogous to the standard treatment of the short range $O(N)$ model 
\cite{PhysRevLett.29.917, Gubser:2017vgc, Giombi:2019enr}
\begin{equation} \label{ActionLongRangeSigma}
S= \frac{C}{2}\int d^d x d^d y \dfrac{\phi^I(x) \phi^I (y)}{|x-y|^{d+s}} +  \int d^d x\left(\frac{1}{2} \sigma \phi^I \phi^I (x)-\frac{\sigma^2}{4\lambda}\right).
\end{equation}
The ordinary $1/N$ perturbation theory of the theory for any $s$ may be developed from the above action (where, in the critical limit, one may drop the quadratic term in $\sigma$) by expanding around the translational invariant vacuum state, where all one-point functions vanish and the propagator of $\sigma$ contributes powers of $1/N$ in correlation functions. However, when we consider a correlation function of operators with a large charge $j$, say $O_j=(\phi^1+i\phi^2)^j$, with $j$ being of the same order as $N$, the standard perturbation theory breaks down. This is because $j$ legs in the operators contribute factors of $j\sim N$ to the action. As we will explain in the next section, in the regime where both $N$ and $j$ are large but $\hat{j} = j/N$ is held fixed, there is a new semiclassical saddle where the operator $\sigma$ acquires a non-trivial classical profile. The two-point function of charge $j$ operators may then be expressed in terms of an effective action at this new saddle point, from which one can extract the scaling dimensions of the operators in this large $j$, large $N$ limit. The scaling dimensions may be expressed as $\Delta_j = N h(\hat{j})$ where $h(\hat{j})$ is a non-trivial function of $d, \hat{j}$ and $s$. We find the following analytic expansions at small and large $\hat{j}$ for generic $s$  
\begin{equation}
\begin{split}
\Delta_j &= N \left[ \frac{d - s}{2} \hat{j} + O(\hat{j}^2) \right]\\
\Delta_j &= N \left[ \frac{d + s}{2} \hat{j} + A(d,s) \hat{j}^{\frac{s}{d + s}} +\ldots \right]. 
\end{split}
\label{Deltaj-summary}
\end{equation}
For small $\hat{j}$, this matches the expectation from the ordinary $1/N$ perturbation theory (the term of order $\hat{j}^2$ can also be explicitly compared to standard diagrammatic expansions). At large $\hat{j}$, the leading behavior of the scaling dimensions is still linear (with a different slope), which is strikingly different from the case of the local $O(N)$ models, where one finds $\Delta_j \sim N \hat{j}^{\frac{d}{d-1}}$ for $\hat{j}\gg 1$ \cite{Alvarez-Gaume:2019biu, Giombi:2020enj}.\footnote{The behavior of the scaling dimension $\Delta_j \sim j^{d/(d-1)}$ in the large charge limit holds in a generic strongly coupled CFT \cite{Hellerman:2015nra, Monin:2016jmo}.} Note that at large $N$, the upper critical value for the range of $s$ is $s^*=2+O(1/N)$. While the behavior (\ref{Deltaj-summary}) arises from the dominant saddle point at generic $s$, we suggest that in the limit $s\rightarrow 2$, the $\Delta_j \sim N \hat{j}^{\frac{d}{d-1}}$ behavior of the short range model is recovered due to an interplay between the multiple solutions to the saddle point equation.  

An interesting aspect of the long range models (\ref{ActionLongRangeQuadratic}) is their connection to the subject of defect CFT. Indeed, the model (\ref{ActionLongRangeQuadratic}) may be thought of as arising from a free scalar field theory in an auxiliary space of dimension $D=d+2-s$, with the quartic interaction localized on a $d$ dimensional ``defect" subspace \cite{Caffarelli,Paulos:2015jfa, Giombi:2019enr}. The operators in the long range $O(N)$ model map to the operators living on the $d$-dimensional defect.
In the special case $s = 1$, i.e. $D=d+1$, the model is equivalent to a BCFT that is free in the bulk and has boundary localized interactions.  In that context, the large charge operators living on the boundary were recently considered in \cite{Cuomo:2021cnb}. They used a Weyl transformation to map the problem of calculating scaling dimensions on the half-plane to that of calculating energies on $R\times HS^{d}$, where $HS^{d}$ is the hemisphere (with the long range model living on the $R\times S^{d-1}$ boundary). To do the calculation, one then may compute the free energy on $R\times HS^{d}$ in the presence of a chemical $\mu$ for the conserved charge. In this paper we generalize this calculation to arbitrary $s$, by mapping the problem to the higher dimensional cylinder $R \times S^{D-1}$ in the presence of a chemical potential $\mu$, with the interaction localized on the subspace $R\times S^{d-1}$. 
Using this approach, we rederive the scaling dimensions of the large charge operators, obtaining results that precisely match with what we get from the saddle point on $R^d$, thus providing a useful consistency check. 
Along the way, we also do perturbative calculations for $s = (d + \epsilon)/2$ in an $\epsilon$ expansion valid for any $N$. We find agreement between large $N$ and $\epsilon$ expansions in the overlapping regimes of validity.

The rest of this paper is organized as follows: In section \ref{Sec:ONFlat}, we set up the calculation of the two-point function of large charge operators in flat space and identify the saddle that provides the dominant contribution in the large $j$, large $N$ double scaling limit. 
Solving the saddle point equation requires calculating the Green's function at the large charge saddle point. Having obtained the Green's function, we show that correlation functions of two ``heavy" (large charge) and an arbitrary number of light operators can be obtained with little extra effort. We discuss in some detail the calculation of ``heavy-heavy-light" three point function and ``heavy-heavy-light-light" four-point function.  We end the section with a separate discussion of the $d = 1$ long range $O(N)$ model that behaves somewhat differently from its higher dimensional counterparts. Then in section \ref{Sec:Cylinder}, we show how our results may be obtained by calculating the energy in a large charge state on a cylinder. The two appendices contain technical details and the standard $1/N$ perturbation theory calculation of the scaling dimensions which is valid at $\hat{j} \ll 1$. 

\section{The large charge, large $N$ saddle point on $R^d$} \label{Sec:ONFlat}
In this section, we start by defining the setup and identifying the large charge saddle point in the long range $O(N)$ model at large $N$. We will closely follow the discussion in \cite{Giombi:2020enj} where the large charge saddle for the local, short-range $O(N)$ models was discussed. Since there are many similarities in the analysis, we will be brief here, and refer the reader to \cite{Giombi:2020enj} for details. 

We will start with the model defined by \eqref{ActionLongRangeSigma}. As mentioned in the introduction, we will study operators that carry a large charge $j$ under $O(N)$ symmetry, and we will work in the double scaling limit such that both $j$ and $N$ are large, with the ratio $\hat{j} = j/ N$ fixed and finite. In this note, we focus on operators that transform in the symmetric traceless representation of $O(N)$. Such operators may be written as $\co_j (x)  \equiv (u^I \phi^I (x))^j$ with null auxiliary complex vector $u^I$ (a simple representative is the operator $(\phi^1+i \phi^2)^j$).  These are the lowest dimension operators in the given charge sector, and are not expected to undergo mixing. Their two-point function is constrained by conformal symmetry in the usual way:
\beq
\braket{\co_j (x_1) \co_j (x_2)} =(u_1^I u_2^I )^j\dfrac{C_j}{x_{12}^{2\Delta_j}}.
\eeq
This two-point function can also be computed using the path integral (recall that in the critical limit we drop the $\sigma$ quadratic term) 
\begin{equation} \label{TwoPointExact}
\begin{split}
\braket{\co_j (x_1) \co_j (x_2)} &= \frac{1}{Z}\int \cD \phi \cD \sigma \left( u_1^I \phi^I (x_1) \right)^j \left( u_2^J \phi^J (x_2) \right)^j e^{-\tfrac{C}{2} \int d^d y d^d z \tfrac{\phi^K(y) \phi^K (z)}{|y-z|^{d+s}} - \frac{1}{2} \int d^d x \sigma \phi^K \phi^K (x)} \\
&=  (u_1^I u_2^I )^j j! \int  \cD \sigma   [G(x_1,x_2;\sigma)]^j e^{-\frac{N}{2} \log\det\lr{\frac{C}{|x-y|^{d+s}}+\sigma(x) \delta^d(x-y))}} \\
& =  (u_1^I u_2^I )^j j! \int  \cD \sigma   e^{-N\{\frac{1}{2}\log\det[\frac{C}{|x-y|^{d+s}}+\sigma(x) \delta^d(x-y)]-\hat{j}\log(G(x_1,x_2;\sigma))\}}
\end{split}
\end{equation}
where we integrated out the scalars and defined the Green's function in the presence of a non-trivial $\sigma$ field
\ba
\delta^{I J} G(x_1,x_2;\sigma) &\equiv \int \cD\phi  \phi^I (x_1) \phi^J (x_2) e^{-\tfrac{C}{2} \int d^d y d^d z \tfrac{\phi^K(y) \phi^K (z)}{|y-z|^{d+s}} - \frac{1}{2} \int d^d x \sigma \phi^K \phi^K (x)}.
\ea
We can perform the path integral over $\sigma$ in \eqref{TwoPointExact} using a saddle point approximation by extremizing the effective action
\beq\label{eqn:saddle}
\dfrac{\delta}{\delta \sigma} \lr{\frac{1}{2}\log\det\left[\frac{C}{|x-y|^{d+s}}+\sigma(x) \delta^d(x-y)\right]-\hat{j}\log(G(x_1,x_2;\sigma))}= 0.
\eeq
This equation will give a profile of $\sigma (x) = \sigma_{*} (x; x_1, x_2)$ satisfying
\beq \label{SaddlePointEqG}
2 \hat{j} G(x_1,x;\sigma_*)G(x_2,x;\sigma_*) = -G(x,x;\sigma_*)G(x_1,x_2;\sigma_*).
\eeq
To calculate the Green's function and then to solve the saddle point equation, we start with an ansatz for the $\sigma$ profile at the saddle point. We observe that this can be viewed as the one-point function of the field $\sigma(x)$ in the presence of the large charge operators, namely a 3-point function:
\begin{equation}
\begin{split}
\sigma_*(x; x_1,x_2) &= \lim_{N\to \infty} \dfrac{\int\cD\sigma \ \sigma(x) e^{-N\{\Seff\}}}{\int \cD\sigma e^{-N\{\Seff\}}}\\
&=\lim_{N\to \infty}  \dfrac{\braket{\co_j(x_1,u_1)\co_j (x_2,u_2)\sigma(x)}}{\braket{\co_j(x_1,u_1) \co_j(x_2,u_2)}}.
\end{split}
\end{equation}
Conformal symmetry then requires
\beq \label{SigmaOnePFLat}
\sigma_* (x; x_1,x_2) = c_\sigma \dfrac{|x_1-x_2|^s}{|x_1-x|^s |x_2-x|^s}
\eeq
since $\Delta_\sigma = s + \co(1/N)$.

\subsection{Green's function} 
In this subsection, we calculate the Green's function in the presence of a non-trivial $\sigma^*$. As usual, it is given by inverting the quadratic term in the action
\beq 
\int d^d x' \lr{\dfrac{C}{|x-x'|^{d+s}} + \sigma_*(x)\delta^d(x-x')} G(x',y;\sigma_*) = \delta^d (x-y)
\eeq. 
We can solve it by expanding the Green's function in powers of $\sigma^*$
\begin{equation}
G = G^{(0)} + G^{(1)} + G^{(2)} + ... 
\end{equation}
where the superscript indicates the power of $\sigma_*$, and the terms in the expansion satisfy following equations 
\begin{equation}
\begin{split}
&\int d^d x' \frac{C}{|x- x'|^{d+s}}  G^{(0)}(x', y) = \delta^d (x- y) \\
& \int d^d x' \frac{C}{|x- x'|^{d+s}}  G^{(L + 1)}(x', y; \sigma^*) = - \sigma_*(x) G^{(L)}(x, y; \sigma^*) , \hspace{ 1cm } L \geq 1.
\end{split}
\end{equation}
The leading order result is just the usual two-point function without any large charge operators 
\begin{equation}
G^{(0)}(x, y) = \frac{C_{\phi}}{|x - y|^{d-s}}, \hspace{1cm} C_{\phi} = \frac{\Gamma \left( \frac{d - s}{2} \right)}{2^s \pi^{\frac{d}{2}} \Gamma \left( \frac{s}{2} \right)}.
\end{equation}
\begin{figure} 
\centering
\includegraphics[scale=0.95]{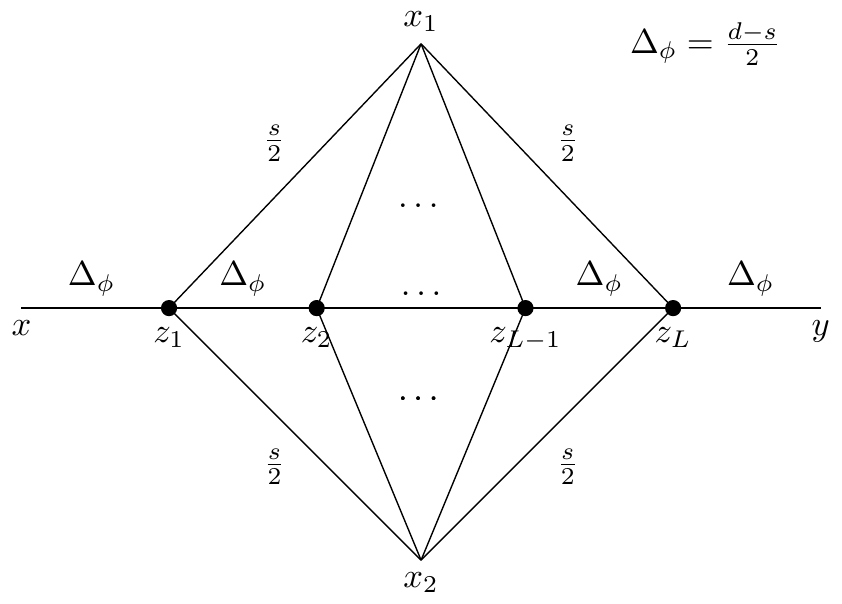}
\caption{The pictorial representation of the integral $I_{L}$. The notation is such that the line between the points $z_i$ and $z_j$ contributes $1/|z_{ij}|^{2 \alpha}$ to the integral where $\alpha$ is the number written above the line. Every filled dot represents a point that is integrated over.}
\label{FigureIntegral}
\end{figure}
We can then get the result for order $L$ Green's function by iteratively applying the above result
\begin{equation} 
\begin{split}
G^{L} (x,y , \sigma_*) &= (-1)^L \left( \prod_{k = 1}^L \int d^d z_k \sigma^* (z_k) G^{0} (z_{k}, z_{k + 1}) \right) G^0 (x, z_1)  \\
&= C_{\phi} I_{L} (x,y, x_1, x_2) \\
I_{L} (x,y, x_1, x_2) &= g_{12}^L  \left( \prod_{k = 1}^L \int \frac{d^d z_k }{|z_k - x_1|^s |z_k - x_2|^s } \right) \left( \prod_{j = 0}^L \frac{1}{|z_{j + 1} - z_j|^{d-s}} \right)
\end{split}
\end{equation}
where we defined $g_{12} = - C_{\phi} c_{\sigma} |x_1 - x_2|^s$ and $z_0 = y, z_{L + 1} = x $. The order $L$ result involves doing an integral over $L$ variables. The integral may be visualized as in figure \ref{FigureIntegral}. Let us now analyze it in more detail. By just shifting all the integration variables, we can see that it is only a function of 3 variables 
\begin{equation}
\begin{split}
I_{L} (x,y, x_1, x_2) = I_{L} (x- x_1,y -x_1, 0, x_2 - x_1). \end{split}
\end{equation}
We can then do a change of variables to invert all the variables $z_k' = z_k/z^2 $, and the fact that the integral is conformal helps to simplify it as follows
\begin{equation}
\begin{split}
&I_{L} (x- x_1,y -x_1, 0, x_2 - x_1) = g_{12}^L  \left( \prod_{k = 1}^L \int \frac{d^d z_k }{|z_k|^s |z_k - (x_2 - x_1)|^s } \right) \left( \prod_{j = 0}^L \frac{1}{|z_{j + 1} - z_j|^{d-s}} \right) \\
&= \frac{1}{|x - x_1|^{d - s} |y - x_1|^{d-s}} \left( \frac{g_{12}}{|x_2 - x_1|^s} \right)^L \left( \prod_{k = 1}^L \int \frac{d^d z'_k }{ |z'_k - (x_2 - x_1)'|^s } \right) \left( \prod_{j = 0}^L \frac{1}{|z'_{j + 1} - z'_j|^{d-s}} \right) \\
& =   \frac{1}{|x - x_1|^{d - s} |y - x_1|^{d-s}} \left( \frac{g_{12}}{|x_2 - x_1|^s} \right)^L \left( \prod_{k = 1}^L \int \frac{d^d z'_k }{ |z'_k |^s } \right) \left( \prod_{j = 0}^L \frac{1}{|z'_{j + 1} - z'_j|^{d-s}} \right)\\
& =   \frac{1}{|x - x_1|^{d - s} |y - x_1|^{d-s} } \left( \frac{g_{12}}{|x_2 - x_1|^s} \right)^L \Phi_{L} (\xi, \eta)
\end{split}
\end{equation}
where in the third line, we did a shift of variables so that in that equation
\begin{equation}
z'_{0} = \eta = \frac{y - x_1}{|y - x_1|^2} - \frac{x_2 - x_1}{|x_2 - x_1|^2}, \hspace{0.5 cm} z'_{L + 1} = \xi = \frac{x - x_1}{|x - x_1|^2} - \frac{x_2 - x_1}{|x_2 - x_1|^2} 
\end{equation}
and in the last line, we defined the integral 
\begin{equation} \label{IntegralGreensFunction}
\Phi_{L} (\xi, \eta) = \left( \prod_{k = 1}^L \int \frac{d^d z'_k }{ |z'_k |^s } \right) \left( \prod_{j = 0}^L \frac{1}{|z'_{j + 1} - z'_j|^{d-s}} \right)
\end{equation} 
\subsubsection*{L = 1}
To gain some intuition for the result, let us start by working perturbatively in $c_{\sigma}$ which is the same as working order by order in $L$. For $L = 1$, we have the following integral 
\begin{equation}
\Phi_{1} (\xi, \eta) =    \int \frac{d^d z }{ |z|^s  |z - \xi|^{d-s} |z - \eta|^{d-s}}. 
\end{equation}
For the purposes of obtaining the scaling dimensions, we will need several limits of the Green's function : $ G(x_1,y, \sigma^*), G(x,x_2, \sigma^*), G(x_1, x_2, \sigma^*)$, and $G(x,x, \sigma^*)$. Let us start with the first one: when $x \rightarrow x_1$. We introduce a regulator $\delta$ and set $x = x_1 + \delta $ to get
\begin{equation}
\Phi_{1} (\xi, \eta) = \delta^{d - s} \int \frac{d^d z }{ |z|^s  |z - \eta|^{d-s}}
\end{equation}
This integral still has a UV divergence, so we introduce a further regulator
\begin{equation} \label{PertGreensLim1}
\begin{split}
\Phi_{1} (\xi, \eta) &=    \delta^{d - s - 2 \kappa} \int \frac{d^d z }{ |z|^{s + \kappa}  |z - \eta|^{d-s + \kappa}}  = \frac{ \delta^{d - s} \pi^{d/2} }{\Gamma \left( \frac{d}{2} \right) \kappa} -  \frac{2  \delta^{d - s} \pi^{d/2} }{\Gamma \left( \frac{d}{2} \right)} \log \left( \delta|\eta| \right) \implies   \\
 G(x_1, y , \sigma^*)  &=  \frac{\Gamma \left( \frac{d - s}{2} \right)}{2^s \pi^{\frac{d}{2}} \Gamma \left( \frac{s}{2} \right) |x_1 - y|^{d-s} } + \frac{\Gamma \left( \frac{d - s}{2} \right)^2 c_{\sigma}}{2^{2 s - 1} \pi^{\frac{d}{2}} \Gamma \left( \frac{s}{2} \right)^2 \Gamma \left( \frac{d}{2} \right) |x_1 - y|^{d-s} } \log \frac{\delta |x_2 - y|}{|x_1 - x_2| |x_1 - y|}.
\end{split}
\end{equation} 
The divergent $1/\kappa$ piece should contribute to wavefunction renormalization, but it will not affect our calculation of scaling dimensions. The $G(x, x_2, \sigma^*)$ should just be related to this one by interchanging $x \leftrightarrow y, x_1 \leftrightarrow x_2$. Finally, to obtain $G(x_1, x_2, \sigma^*)$, we set $y = x_2 + \delta$ in the above to get 
\begin{equation} \label{PertGreensLim2}
\begin{split}
G(x_1, y , \sigma^*)  =  \frac{\Gamma \left( \frac{d - s}{2} \right)}{2^s \pi^{\frac{d}{2}} \Gamma \left( \frac{s}{2} \right) |x_1 - x_2|^{d-s} } + \frac{\Gamma \left( \frac{d - s}{2} \right)^2 c_{\sigma}}{2^{2 s - 1} \pi^{\frac{d}{2}} \Gamma \left( \frac{s}{2} \right)^2 \Gamma \left( \frac{d}{2} \right) |x_1 - x_2|^{d-s} } \log \frac{\delta^2 }{|x_1 - x_2|^2}.
\end{split}
\end{equation}

Next we turn to $G(x,x, \sigma^*)$, for which we need to consider $\xi = \eta$ limit of the integral \eqref{IntegralGreensFunction} which is easy to obtain 
\begin{equation}
\Phi_{1} (\xi, \xi) = \frac{\pi^{d/2} \Gamma \left(\frac{d - s}{2} \right)^2 \Gamma \left( s - \frac{d}{2} \right) }{\Gamma \left(d - s \right) \Gamma \left(\frac{s}{2} \right)^2  |\xi|^{d-s}}
\end{equation}
which implies 
\begin{equation} \label{PerturbativeGreensCoinc}
G^{(1)} (x,x, \sigma_*) = - \frac{ C_{\phi}^2 c_{\sigma} \pi^{d/2} \Gamma \left(\frac{d - s}{2} \right)^2 \Gamma \left( s - \frac{d}{2} \right) |x_2 - x_1|^{d-s} }{\Gamma \left(d - s \right) \Gamma \left(\frac{s}{2} \right)^2  |x - x_1|^{d-s} |x - x_2|^{d-s}}.
\end{equation}
Note that without any insertions, we should set coincident point two-point function to zero, i.e. $G^{(0)} (x,x) = 0$.

\subsubsection*{General $L$}
Next we turn to the more difficult problem of evaluating the integral in \eqref{IntegralGreensFunction} for general $L$. For the local case of $s = 2$, the integral is exactly the kind considered in \cite{Isaev:2003tk}. But the generalization of that formalism to general $s$ is not straightforward. Fortunately, the integral in \eqref{IntegralGreensFunction} was already computed exactly in \cite{Derkachov:2021ufp} in a very different context of fishnet Feynman integrals. The result can be expressed in terms of Gegenbauer polynomials as follows 
\begin{equation} \label{IntegralResult}
\begin{split}
&\Phi_{L} (x, y) = \frac{\Gamma \left( \frac{d - 2}{2} \right)  }{ (|x| |y|)^{\frac{d - s}{2}} \pi^{\frac{-d L}{2}}} \sum_{l = 0}^{\infty} \left( l + \frac{d- 2}{2} \right)  C_{l}^{ \left( \frac{d- 2}{2} \right)} \left(\frac{x \cdot y}{|x||y|} \right) \int \frac{d u}{2 \pi} \left( \frac{x^2}{y^2} \right)^{i u} \left( Q_{l}(u) \right)^{L + 1} \\
& Q_{l} (u) = \frac{\Gamma \left(\frac{s}{2}\right) \Gamma \left(\frac{d - s + 2 l}{4} - i u \right) \Gamma \left(\frac{d - s + 2 l }{4} + i u\right) }{\Gamma \left( \frac{d - s}{2} \right) \Gamma \left(\frac{d + s + 2 l}{4} + i u \right) \Gamma \left(\frac{d + s + 2 l}{4} - i u\right)}
\end{split}
\end{equation}
In principle, the integral over $u$ can be performed by a sum over residues. The contour can be closed in the upper half-plane and there are infinitely many residues at $u = i \left( \frac{d - s + 2 l}{4} + n\right)$. But in practice, it becomes hard since the poles are of order $L + 1$. To convince the reader that it makes sense, we perform some basic checks of this result starting from $L = 0$ when there is no integral to do. This can be expressed in terms of Gegenbauer polynomials as follows
\begin{equation}
\Phi_{0} (x, y) = \frac{1}{|x - y|^{d - s}} = \frac{1}{|x|^{d - s}} \sum_{\alpha = 0}^{\infty} C_{\alpha}^{\frac{d - s}{2}} \left(\frac{x \cdot y}{|x||y|}  \right) \left| \frac{y}{x} \right|^{\alpha}
\end{equation}
On the other hand, the integral on the right hand side of \eqref{IntegralResult} gives the following sum over residues 
\begin{equation}
\Phi_{0} (x, y) = \frac{ \Gamma \left( \frac{d - 2}{2} \right)  }{ |x|^{d - s} \Gamma \left( \frac{d - s}{2} \right) } \sum_{l, n = 0}^{\infty}  \frac{ \left( l + \frac{d- 2}{2} \right)  \Gamma \left( \frac{d - s}{2} + l + n \right) \Gamma \left( 1 + n - \frac{s}{2} \right) }{ n! \Gamma \left( \frac{d}{2} + l + n \right) \Gamma \left( 1 - \frac{s}{2} \right)} C_{l}^{  \frac{d- 2}{2} } \left(\frac{x \cdot y}{|x||y|} \right) \left| \frac{y}{x} \right|^{l + 2 n}
\end{equation}
This implies the following identity for Gegenbauer polynomials 
\begin{equation} \label{GegenbauerIdentity}
C_{\alpha}^{\frac{d - s}{2}} (x) = \sum_{n = 0}^{\lfloor \frac{\alpha}{2} \rfloor} C_{\alpha - 2 n}^{  \frac{d- 2}{2}} (x)   \frac{ \left( \frac{d- 2}{2} + \alpha - 2 n \right)  \Gamma \left( \frac{d - s}{2} + \alpha - n \right) \Gamma \left( 1 + n - \frac{s}{2} \right) \Gamma \left( \frac{d - 2}{2} \right) }{ n! \Gamma \left( \frac{d}{2} + \alpha - n \right) \Gamma \left( 1 - \frac{s}{2} \right) \Gamma \left( \frac{d - s}{2} \right) } . 
\end{equation}
which is known to be true (see for instance Appendix A of \cite{Kotikov:2018wxe}). 

We can get the full Green's function by summing over $L$
\begin{equation} \label{GreensFunctionSum}
\begin{split}
G (x,y, \sigma^*) = \sum_{L = 0}^{\infty} G^L =& \frac{C_{\phi} \Gamma \left( \frac{d - 2}{2} \right)}{|x - x_1|^{d - s} |y - x_1|^{d-s} \left( |\xi| |\eta| \right)^{\frac{d - s}{2}}} \sum_{l = 0}^{\infty} \left( l + \frac{d- 2}{2} \right)  C_{l}^{ \left( \frac{d- 2}{2} \right)} \left(\frac{\xi \cdot \eta}{|\xi||\eta|} \right) \\
& \times \int \frac{d u}{2 \pi} \left( \frac{\xi^2}{\eta^2} \right)^{i u} \frac{Q_{l} (u)}{1 + C_{\phi} c_{
\sigma} \pi^{d/2} Q_{l} (u)}.
\end{split}
\end{equation}
It is useful to write this Green's function as a function of the conformal cross-ratios defined as 
\begin{equation}
X = \frac{|x - x_1|^2 |y - x_2|^2}{|x_1 - x_2|^2 |x - y|^2}, \hspace{1cm} Y = \frac{|x - x_2|^2 |y - x_1|^2}{|x_1 - x_2|^2 |x - y|^2}.
\end{equation}
The Green's function in terms of these cross-ratios is given by 
\begin{equation} \label{GreensFunctionCrossRatio}
\begin{split}
G (x,y, \sigma^*) =  & \frac{C_{\phi} \Gamma \left( \frac{d - 2}{2} \right)}{|x - y|^{d - s} \left( X Y \right)^{\frac{d - s}{4}}} \sum_{l = 0}^{\infty} \left( l + \frac{d- 2}{2} \right)  C_{l}^{ \left( \frac{d- 2}{2} \right)} \left(\frac{X + Y - 1}{2 \sqrt{X Y}} \right) \\
& \times \int \frac{d u}{2 \pi} \left( \frac{Y}{X} \right)^{i u} \frac{Q_{l} (u)}{1 + C_{\phi} c_{
\sigma} \pi^{d/2} Q_{l} (u)}
\end{split}
\end{equation}
The reason that the Green's function can be written in terms of the standard conformal cross-ratios is that, as we clarify below, it is directly related to the 4-point function of two large charge and two charge 1 operators. 

One may perform the integral over $u$ by closing the contour in the upper half plane, and the poles are given by the solutions to the following equation
\begin{equation} \label{EqWCSigma}
\frac{1}{Q_l(u)} + C_{\phi} c_{
\sigma} \pi^{d/2} = 0, \implies  \frac{ \Gamma \left(\frac{d + s + 2 l}{4} - i u \right) \Gamma \left(\frac{d + s + 2 l }{4} + i u\right) }{\Gamma \left(\frac{d - s + 2 l}{4} + i u \right) \Gamma \left(\frac{d - s + 2 l}{4} - i u\right)} = - \frac{c_{\sigma}}{2^s}.
\end{equation}
We expect the poles to lie on the imaginary axis, and we will parameterize the roots of the above equation by $u=i\mu/2$, where $\mu=\mu(c_{\sigma})$ is real.\footnote{We denote it by $\mu$ because it will be equal to the chemical potential when we map the problem to the cylinder.} 

We can now calculate this Green's function in the various limits needed to extract the scaling dimensions. Let us start by considering the case when either $x\rightarrow x_1$ or $ y \rightarrow x_2 $ or both at the same time. In the limit when $x \rightarrow x_1$, we have $\xi \rightarrow \infty$, the dominant contribution to the integral in \eqref{GreensFunctionSum} comes from the pole with the smallest positive imaginary part, i.e. $u=i\mu/2$ with the smallest $\mu$. This also allows to just consider $l = 0$ term in the sum. The same is true when $y \rightarrow x_2$, because in that case, $\eta \rightarrow 0$. 
We can find this solution analytically for small and large $c_{\sigma}$ as 
\begin{equation} \label{RootLimits}
\begin{split}
&\mu(c_{\sigma}) = \frac{d - s}{2} + \frac{ \Gamma \left(\frac{d - s}{2} \right) c_{\sigma}  }{ 2^{s - 1}\Gamma \left(\frac{d}{2} \right)\Gamma \left(\frac{s}{2} \right)} + \frac{\Gamma \left(\frac{d-s}{2}\right)^2 \left(\psi ^{(0)}\left(\frac{d-s}{2}\right)-\psi ^{(0)}\left(\frac{d}{2}\right)+\psi ^{(0)}\left(\frac{s}{2}\right)+\gamma \right) c_{\sigma}^2}{2^{2 s - 1} \Gamma \left(\frac{d}{2}\right)^2 \Gamma \left(\frac{s}{2}\right)^2} + \dots \\
&\mu(c_{\sigma}) =  \frac{d + s}{2} + \frac{2^{s + 1} \Gamma \left(\frac{d + s}{2} \right)}{ \Gamma \left(\frac{d}{2} \right)\Gamma \left( \frac{-s}{2} \right) c_{\sigma}} + \frac{\Gamma \left(\frac{d+s}{2}\right)^2 \left(\psi ^{(0)}\left(\frac{d+s}{2}\right)-\psi ^{(0)}\left(\frac{d}{2}\right)+\psi ^{(0)}\left(-\frac{s}{2}\right)+\gamma \right)}{2^{-1-2 s}  \Gamma \left(\frac{d}{2}\right)^2 \Gamma \left(-\frac{s}{2}\right)^2 c_{\sigma}^2} + \dots
\end{split}
\end{equation}
For general values of $c_{\sigma}$, we can find this root numerically. The Green's function in this limit is then given by 
\begin{equation} \label{FullGreenLim}
\begin{split}
&G (x_1,y, \sigma^*) =  \frac{\Gamma \left(\frac{d }{2} \right)}{2 \pi^{d/2} |y - x_1|^{d-s}} \mu'(c_{\sigma}) \left( \frac{\delta |y - x_2| }{|x_2 - x_1||y - x_1|} \right)^{\mu(c_{\sigma}) - \frac{d - s}{2}} \\
& G (x,x_2, \sigma^*) =  \frac{\Gamma \left(\frac{d }{2} \right)}{2 \pi^{d/2} |x - x_2|^{d-s}} \mu'(c_{\sigma}) \left( \frac{\delta |x - x_1| }{|x_2 - x_1||x - x_2|} \right)^{\mu(c_{\sigma}) - \frac{d - s}{2}} \\ 
&G (x_1,x_2, \sigma^*) =  \frac{\Gamma \left(\frac{d }{2} \right)}{2 \pi^{d/2} |x_2 - x_1|^{d-s}} \mu'(c_{\sigma}) \left( \frac{\delta^2 }{|x_2 - x_1|^2} \right)^{\mu(c_{\sigma}) - \frac{d - s}{2}} \\
\end{split}
\end{equation}
At small $c_{\sigma}$, this can be checked to agree with \eqref{PertGreensLim1} and \eqref{PertGreensLim2}.

Finally, we consider the coincident point limit, $G(x,x, \sigma^*)$. In this limit, $\xi \rightarrow \eta$ and the result simplifies to 
\begin{equation} \label{CoincidentPointGreens}
\begin{split}
G^L(x,x, \sigma^*) &= \frac{C_{\phi} \Gamma \left( \frac{d - 2}{2} \right)|x_2 - x_1|^{d - s} \left( - C_{\phi} c_{
\sigma} \pi^{d/2} \right)^L }{\left( |x - x _1| |x - x_2| \right)^{d-s}} \\
& \times \sum_{l = 0}^{\infty} \left(\frac{2 l + d- 2}{2} \right)  C_{l}^{ \left( \frac{d- 2}{2} \right)} \left(1 \right) \int \frac{d u}{2 \pi}  \left( Q_{l} (u) \right)^{L + 1}.
\end{split}
\end{equation}
As usual, this coincident point limit is related to the functional determinant of the quadratic piece in the action, so we discuss it more in the next subsection.

\subsection{Functional determinant}
The functional determinant can be expressed in terms of the Green's function as follows
\begin{equation} \label{FunctDetDef}
\begin{split}
\log \det \left[\frac{C}{|x-y|^{d+s}}+\sigma_*(x) \delta^d(x-y)\right] &= \sum_{L = 1}^{\infty}  \frac{(-1)^{L - 1}}{L} \left( \prod_{i = 1}^{L} \int  d^d z_i \sigma_*(z_i) G^0 (z_i, z_{i + 1}) \right) \\
&= \sum_{L = 1}^{\infty} \frac{1}{L} \int d^d x \sigma_*(x) G^{L - 1} (x,x, \sigma_*)
\end{split}
\end{equation}
where in the first line, it is to be understood that $z_{L + 1} = z_1$. Plugging the result from \eqref{CoincidentPointGreens}, we need to perform an integral over $x$, which is divergent. We regularize it in the same way as we did in previous subsection 
\begin{equation} \label{DivIntegral}
\begin{split}
\int d^d x \frac{|x_2 - x_1|^{d}}{|x - x_1|^d |x - x_2|^{d}} &\rightarrow \int d^d x \frac{ \delta^{-2 \kappa} |x_2 - x_1|^{d - 2 \kappa}}{|x - x_1|^{d - 2 \kappa} |x - x_2|^{d - 2 \kappa}}   \\
&= \frac{2 \pi^{d/2}}{\kappa  \Gamma \left( \frac{d}{2} \right)} - \frac{4 \pi^{d/2}}{\Gamma \left( \frac{d}{2} \right)} \log \left( \frac{\delta}{ |x_1 - x_2|} \right) + O(\kappa).
\end{split}
\end{equation}
Again, the $1/ \kappa$ piece will be canceled by an appropriate counterterm and will not be important for us, so we will only keep the $\log$ term in the following. Then performing the sum over $L$, we obtain the following result for the functional determinant 
\begin{equation} \label{FunctDetRes}
\begin{split}
&\log \det [...] = - 2 F (c_{\sigma}) \log \left( \frac{\delta^2}{ |x_{12}|^2} \right) \\
& F(c_{\sigma}) =  \sum_{l = 0}^{\infty} \frac{\left(2 l + d - 2 \right) \Gamma \left( d - 2 + l\right)}{\Gamma \left( d - 1 \right) l! }  \int \frac{d u}{2 \pi}  \log \left( 1 + C_{\phi} c_{\sigma} \pi^{d/2} Q_{l} (u) \right) 
\end{split}
\end{equation}
where we used 
\begin{equation} \label{Gegenbauer1}
C_{l}^{ \left( \frac{d- 2}{2} \right)} \left(1 \right) = \frac{\Gamma \left( d- 2 + l\right)}{\Gamma \left( d- 2 \right) \Gamma \left( l  + 1 \right)}.
\end{equation}
The Green's function at coincident points is related to the derivative of the functional determinant. Indeed by summing over $L$ in \eqref{CoincidentPointGreens}, we can see
\begin{equation} \label{CoincPointRes}
\begin{split}
G(x,x, \sigma^*) &= \sum_{l = 0}^{\infty}  \frac{C_{\phi} \Gamma \left( \frac{d - 2}{2} \right)|x_2 - x_1|^{d - s} (2 l + d - 2) \Gamma \left( d - 2 + l \right)}{ 2 \Gamma \left( d - 2 \right)  l! \left( |x - x _1| |x - x_2| \right)^{d-s} } \int \frac{d u}{2 \pi} \frac{Q_l(u)}{\left(1 + C_{\phi} c_{\sigma} \pi^{d/2} Q_l(u) \right)} \\
&= \frac{ \Gamma \left( \frac{d}{2} \right) F'(c_{\sigma})}{\pi^{\frac{d}{2}}} \left(\frac{|x_2 - x_1|}{|x - x _1| |x - x_2|} \right)^{d-s}.
\end{split}
\end{equation}

In the limit of small $c_{\sigma}$, we can expand in powers of $c_{\sigma}$
\begin{equation}
F(c_{\sigma}) = -\sum_{l = 0}^{\infty} \frac{\left(2 l + d - 2 \right) \Gamma \left( d - 2 + l\right)}{  \Gamma \left( d - 1 \right) l! }  \int \frac{d u}{2 \pi} \sum_{L = 1}^{\infty} \frac{\left(- C_{\phi} c_{\sigma} \pi^{d/2} Q_{l} (u) \right)^{L}} {L}.
\end{equation}
For the $L = 1$ term, we can explicitly perform the integral by summing up residues, and then perform the sum over $l$ to check that it vanishes. So in the limit of small $c_{\sigma}$, $F(c_{\sigma})$ actually goes like $c_{\sigma}^2$. This is also expected because the $L = 1$ term is proportional to $G^0 (x,x)$ which is the short distance limit of the two-point function in flat space, which should be set to zero. To obtain this $c_{\sigma}^2$ it is easiest to go back to \eqref{FunctDetDef} and look at the $L = 2$ term. The Green's function $G^1$ was written in \eqref{PerturbativeGreensCoinc} and after performing the integral over $x$, we get 
\begin{equation} \label{FSmallcsigma}
F(c_{\sigma}) = - \frac{ c_{\sigma}^2 \Gamma \left(\frac{d - s}{2} \right)^4 \Gamma \left( s - \frac{d}{2} \right) }{ 2^{2 s + 1} \Gamma \left(d - s \right) \Gamma \left(\frac{s}{2} \right)^4  \Gamma \left(\frac{d}{2} \right) }.
\end{equation} 
As a check, note that in the special case of $ s= 2$, corresponding to the local (short range) $O(N)$ model, the expression of $Q_{l}(u)$ simplifies to
\begin{equation}
Q_{l} (u) = \frac{1}{\Gamma \left( \frac{d - 2}{2}\right) \left( u^2 + \frac{1}{4} \left(l + \frac{d - 2}{2} \right)^2 \right)}.
\end{equation}
We can then perform the integral over $u$ by closing the contour in the upper half plane and using the residue at $u = i \left(\frac{d - 2 + 2 l}{4} \right)$. This gives for $s = 2$
\begin{equation}
F(c_{\sigma}) = \sum_{k = 0}^{\infty}  \frac{(-1)^{k + 1} c_{\sigma}^{k + 2} (2 k + 1)! }{2^{2 k + 2} k! (k + 2)! \Gamma(d-1)} \sum_{l = 0}^{\infty}\frac{ \Gamma \left( d - 2 + l\right)}{   \left(l + \frac{d}{2} - 1 \right)^{2 k + 2} l! } . 
\end{equation}
It can be checked that this agrees with what was obtained in \cite{Giombi:2020enj}.

To find the large $\hat{j}$ limit of the scaling dimensions, we will also need the large $c_{\sigma}$ behavior of $F(c_{\sigma})$. To gain some intuition, note that for a constant mass, we have 
\begin{equation}
\frac{1}{2} \textrm{Tr} \log \left( (-\nabla^2)^{\frac{s}{2}} + m^2 \right) = \frac{\textrm{Vol} (R^d)}{2} \int \frac{d^d p}{(2 \pi)^d} \log \left( p^s + m^2 \right) = \frac{\textrm{Vol} (R^d) (m^2)^{\frac{d}{s}}}{2^{d + 1} \pi^{\frac{d}{2} - 1} \Gamma \left( \frac{d}{2} + 1 \right) \sin \left( \frac{\pi d}{s} \right)}.
\end{equation}
A natural guess in the presence of a position dependent $\sigma^* $ is that we should replace 
\begin{equation}
\begin{split}
\textrm{Vol} (R^d) (m^2)^{\frac{d}{s}} &\rightarrow \int d^d x (\sigma_* (x))^{\frac{d}{s}} \implies \\ 
 \frac{1}{2} \textrm{Tr} \log \left( (-\nabla^2)^{\frac{s}{2}} + \sigma_*(x) \right) &= \frac{\int d^d x (\sigma_* (x))^{\frac{d}{s}}}{2^{d + 1} \pi^{\frac{d}{2} - 1} \Gamma \left( \frac{d}{2} + 1 \right) \sin \left( \frac{\pi d}{s} \right)} \\
& =- \frac{(c_{\sigma})^{\frac{d}{s}} \pi  }{2^{d - 1} \ d  \ \Gamma \left( \frac{d}{2} \right)^2 \sin \left( \frac{\pi d}{s} \right)}\log \left( \frac{\delta^2}{ |x_{12}|^2} \right)
\end{split}
\end{equation}
We will show in appendix \ref{App:largecF} that this is indeed the correct behavior using heat kernel methods. So in the limit of large $c_{\sigma}$, we have \eqref{FcsLargecApp}
\begin{equation} \label{LargecF}
F(c_{\sigma}) =  \frac{(c_{\sigma})^{\frac{d}{s}} \pi  }{2^{d - 1} \ d  \ \Gamma \left( \frac{d}{2} \right)^2 \sin \left( \frac{\pi d}{s} \right)} \left( 1 + O \left( \frac{1}{c_{\sigma}^{2/s}} \right) \right).
\end{equation}

\begin{figure} 
\centering
\includegraphics[scale=0.75]{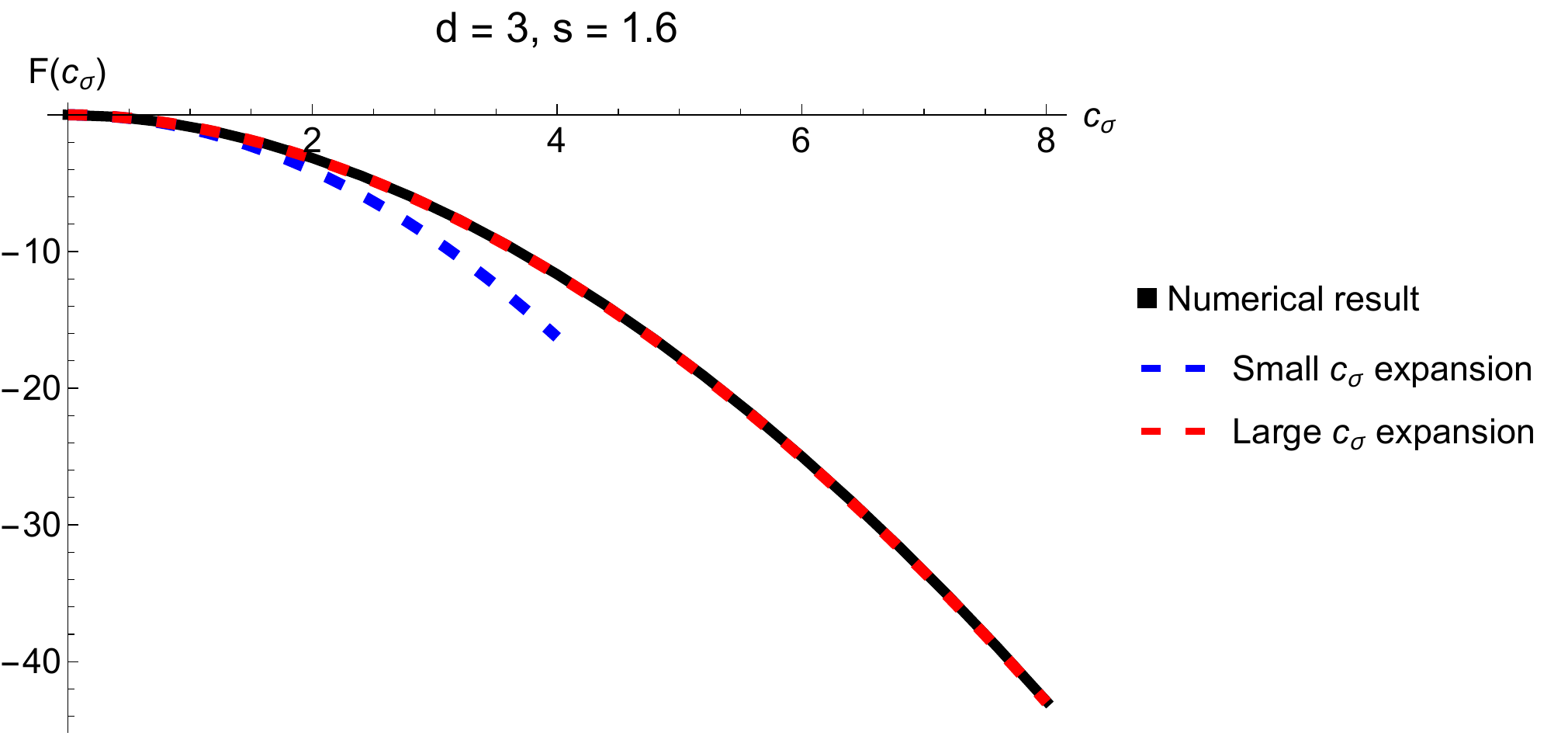}
\caption{The numerical result for $F(c_{\sigma})$ in $d = 3,  s = 1.6$. We also plot the analytic expansions at small and large $c_{\sigma}$.}
\label{FigureFcNumerics}
\end{figure}

For finite $c_{\sigma}$, one can evaluate the functional determinant numerically using \eqref{FunctDetRes}. But for the numerics to converge, it is necessary to regulate it. One simple way to do this is to use the following form
\begin{equation} \label{FcReg}
\begin{split}
F(c_{\sigma}) =& \sum_{l = 0}^{\infty} \frac{\left(2 l + d - 2 \right) \Gamma \left( d - 2 + l\right)}{\Gamma \left( d - 1 \right) l! }  \int_0^{\infty}\frac{ d u}{\pi}  \bigg[ \log \left( 1 + C_{\phi} c_{\sigma} \pi^{d/2} Q_{l} (u) \right) \\
&- C_{\phi} c_{\sigma} \pi^{d/2} Q_{l} (u) + \frac{\left(C_{\phi} c_{\sigma} \pi^{d/2} Q_{l} (u) \right)^2}{2} \bigg]  - \frac{ c_{\sigma}^2 \Gamma \left(\frac{d - s}{2} \right)^4 \Gamma \left( s - \frac{d}{2} \right) }{ 2^{2 s + 1} \Gamma \left(d - s \right) \Gamma \left(\frac{s}{2} \right)^4  \Gamma \left(\frac{d}{2} \right) },
\end{split}
\end{equation} 
where we subtracted out the linear and quadratic pieces in $c_{\sigma}$ and then added them back (the linear in $c_{\sigma}$ term vanishes while the quadratic term is given by \eqref{FSmallcsigma}). To avoid confusion, we emphasize that the last term in the above formula is not integrated over $u$ or summed over $l$. This formula can be directly used to numerically evaluate $F(c_{\sigma})$. We plot the results for $d = 3, s = 1.6$ in \eqref{FigureFcNumerics}. As is clear, the large $c_{\sigma}$ result works very well even down to very small $c_{\sigma}$. 

\subsection{The scaling dimensions}
We finally have all the ingredients to calculate the scaling dimensions of the operators $\mathcal{O}_j$, which can be extracted from the two-point function 
\begin{equation}
\Delta_j = -\frac{1}{2} |x_{12}| \frac{\partial}{\partial |x_{12}| } \log \braket{\co_j (x_1) \co_j (x_2)}.
\end{equation}
At the large $N$ saddle point, using \eqref{TwoPointExact}, this is given by 
\begin{equation} \label{ScalingDimRes}
\begin{split}
\Delta_j &= \frac{N }{2} |x_{12}| \frac{\partial}{\partial |x_{12}| } \lr{\frac{1}{2}\log\det\left[\frac{C}{|x-y|^{d+s}}+\sigma(x) \delta^d(x-y)\right]-\hat{j}\log(G(x_1,x_2;\sigma))} \\
&=  N \left( F(c_{\sigma}) + \hat{j} \mu(c_{\sigma}) \right)
\end{split}
\end{equation}
where we used \eqref{FullGreenLim} and \eqref{FunctDetRes}. The number $c_{\sigma}$ is determined by solving the saddle point equation \eqref{SaddlePointEqG}, which after using \eqref{FullGreenLim} and \eqref{CoincPointRes} becomes 
\begin{equation} \label{SaddleEqRes}
F'(c_{\sigma}) = - \hat{j} \mu'(c_{\sigma}).
\end{equation}
Note that this just corresponds to extremizing $\Delta_j=N(F(c_{\sigma})+\hat{j}\mu(c_{\sigma}))$ with respect to the constant $c_{\sigma}$. 

\subsubsection*{Small $\hat{j}$ expansion }
At small $c_{\sigma}$, we can use \eqref{FSmallcsigma} and $\eqref{RootLimits}$ to get the solution to the saddle point equation 
\begin{equation} \label{SmalljSaddSol}
\begin{split}
c_{\sigma} &= \frac{\hat{j} 2^{s + 1} \Gamma \left(d - s \right) \Gamma \left( \frac{s}{2} \right)^3 }{\Gamma \left( \frac{d - s}{2} \right)^3 \Gamma \left( s - \frac{d}{2} \right) } \implies \\
\frac{\Delta_j }{N}  &=  \frac{d - s}{2} \hat{j} + \frac{2 \Gamma \left(d - s \right) \Gamma \left( \frac{s}{2} \right)^2  }{ \Gamma \left( \frac{d - s}{2} \right)^2 \Gamma \left( s - \frac{d}{2} \right) \Gamma \left( \frac{d}{2} \right) } \hat{j}^2  + O(\hat{j}^3).
\end{split}
\end{equation}
Note that, recalling that $\hat{j}=j/N$, the quadratic term in $\hat{j}$ above should match the term proportional to $j^2$ in the anomalous dimension to order $1/N$ computed in the standard large $N$ diagrammatic expansion. We check this explicitly in appendix \ref{App:Pert1N}.

In the next section, we will also study this model in an $\epsilon$ expansion in $s = \frac{d + \epsilon}{2}$ for any $N$ but with $\epsilon j$ held fixed. To compare with the results in that section, we write here the above result for this value of $s$ to leading order in $\epsilon$ 
\begin{equation} \label{LargeNsmalljEps}
\Delta_j   = j \left(  \frac{d }{4}  + \frac{\epsilon j  }{N} + O \left( (\epsilon j)^2 \right) \right). 
\end{equation}

\subsubsection*{Large $\hat{j}$ expansion}
At large $c_{\sigma}$, using \eqref{LargecF} and \eqref{RootLimits} , the saddle point equation gives 
\begin{equation} \label{LargejSaddSol}
c_{\sigma} = \left( \frac{\hat{j} \ 2^{d + s} \ s  \ \Gamma \left( \frac{d + s}{2} \right)  \Gamma \left( \frac{d}{2} \right) \sin \left( \frac{\pi d}{s} \right) } {\pi  \Gamma \left(- \frac{s}{2} \right) } \right)^{\frac{s}{d + s}} 
\end{equation}
This gives the dimension of the large charge operator in the limit of large $\hat{j}$
\begin{equation}
\begin{split}
\frac{\Delta_j }{N} &= \frac{d + s}{2} \hat{j} + A(d,s) \hat{j}^{\frac{d}{d + s}} \\
A(d,s) &= \frac{2 \pi (d + s)}{\Gamma \left( \frac{d}{2} \right)^2 \sin \left( \frac{\pi d}{s} \right) d s } \left( \frac{ \Gamma \left( \frac{d + s}{2} \right)  s \sin \left( \frac{\pi d}{s} \right) \Gamma \left( \frac{d}{2} \right) }{ \Gamma \left(- \frac{s}{2} \right) \pi } \right) ^{\frac{d}{d + s}} 
\end{split}
\end{equation}
Notice that the factor in the parenthesis above becomes negative for $s > 2$ and for $s < d/2$ (we are considering only $d > 2$ case here). This implies that outside of the range $d/2 < s < 2$, the factor $A(d,s)$ becomes complex. 
This is consistent with the fact that the long range real fixed points only exists in the range  $d/2 < s < s^*$, with $s^*=2+O(1/N)$. As mentioned in the introduction, for $s < d/2$ the IR limit of the long range model is described by the Gaussian fixed point, while for $s \ge 2$ the system should cross over to the short range fixed point. We will comment on this more in subsection \ref{cross}. 

For $s = \frac{d + \epsilon}{2}$, where we can compare to the weakly coupled Wilson-Fisher fixed point, the expansion in $\epsilon j$ of the above result at leading order in $\epsilon$ yields    
\begin{equation}  \label{LargeNlargejEps}
\Delta_j  = j \left( \frac{3 d}{4}  - \frac{3 \left( \sin \left(\frac{\pi  d}{4}\right)  \Gamma \left(\frac{d}{4}+1\right) \Gamma \left(\frac{3 d}{4}\right)\right)^{2/3}}{\pi ^{2/3} \Gamma \left(\frac{d}{2}\right)^{4/3}} \left( \frac{N}{\epsilon j} \right)^{1/3} \right). 
\end{equation}

\begin{figure} 
\centering
\begin{subfigure}{0.48\textwidth}
\includegraphics[width = \textwidth]{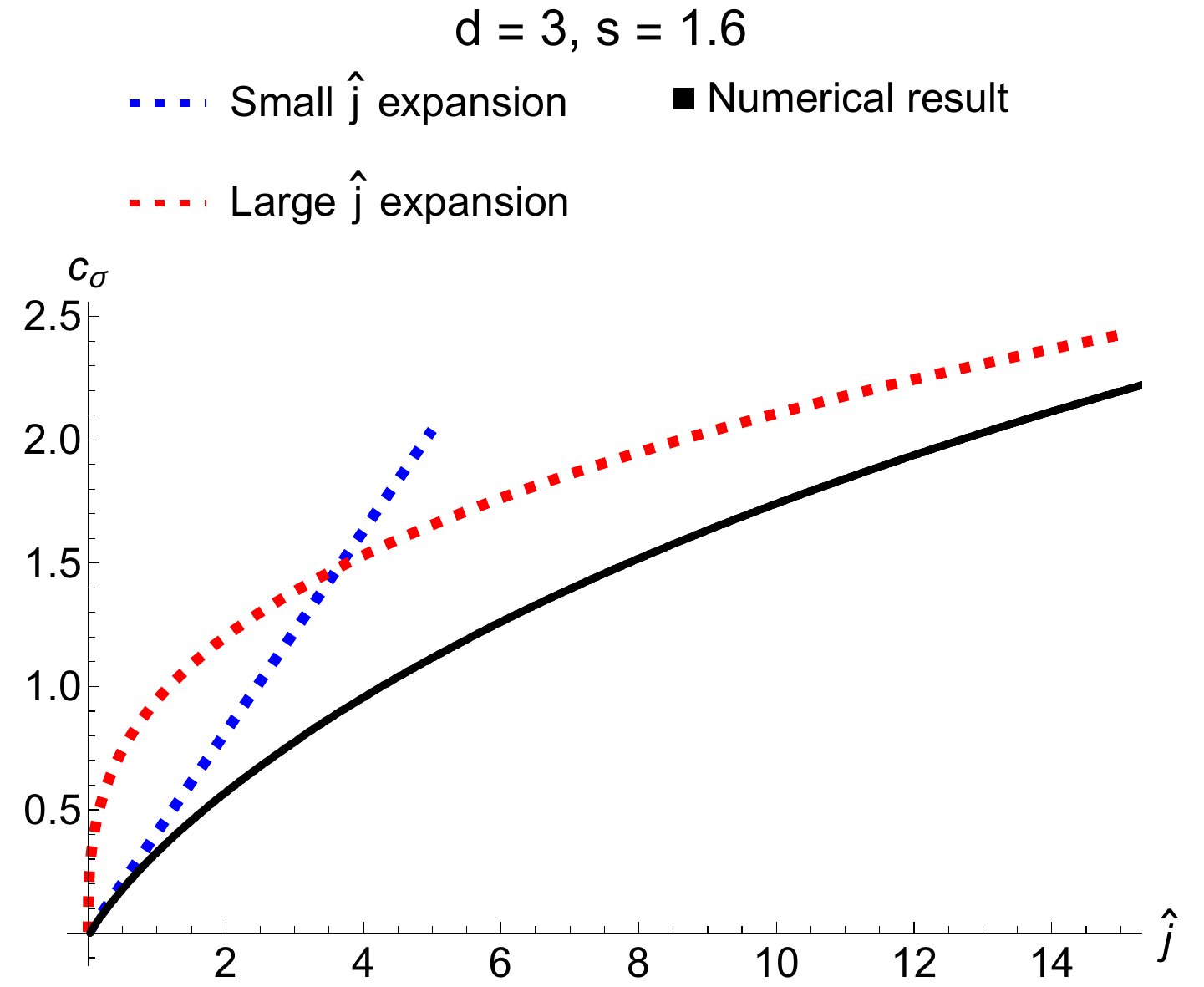}
\end{subfigure}
\begin{subfigure}{0.48\textwidth}
\includegraphics[width = \textwidth]{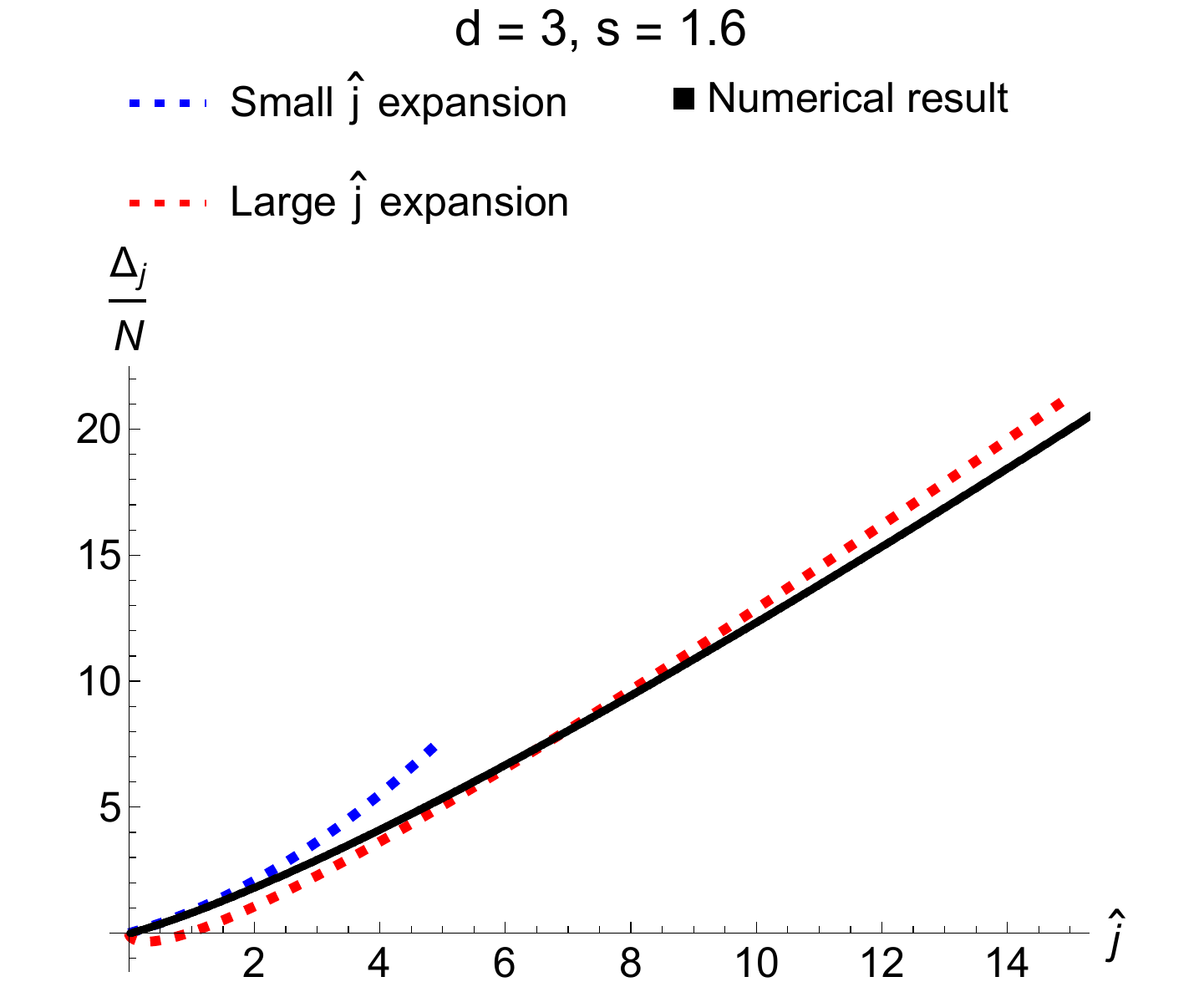}
\end{subfigure}
\caption{The numerical results for the dimension $\Delta_j$ and the solution to the saddle point equation for $d = 3, s = 1.6$. In both the plots, black line represent the numerical results, the dashed red line is the analytical result in a large $\hat{j}$ expansion and the dashed blue line is the analytical result in a small $\hat{j}$ expansion.}
\label{FigureDeltawj16}
\end{figure}

For general intermediate $ \hat{j}$, we can numerically evaluate $F(c_{\sigma})$ and $\mu(c_{\sigma})$ and then solve the saddle point equation \eqref{SaddleEqRes} numerically to obtain $c_{\sigma}$ as a function of $\hat{j}$. We can then plug in this into \eqref{ScalingDimRes} to get the scaling dimensions. We show these numerical results for $d = 3, s = 1.6$ in figure \ref{FigureDeltawj16}.

\subsection{Correlation functions}
So far we have focused on the two-point function of large charge operators. But having access to the Green's function at the large charge saddle point, it is easy to obtain also higher point correlation functions involving two heavy (large charge) and an arbitrary number of light (finite charge) operators. In this subsection, we will focus on the ``heavy-heavy-light" three point function and ``heavy-heavy-light-light" four point function. We will closely follow the approach used in \cite{Giombi:2020enj}. 

Let us start with  the three point function, and as before we will consider scalar operators in symmetric traceless representation of $O(N)$, which may be written as $\mathcal{O}_j = (u \cdot \phi)^j$. Their three-point function is fixed by conformal symmetry and $O(N)$ symmetry upto an overall constant 
\begin{equation}
\begin{split}
&\langle \mathcal{O}_{j_1}(x_1, u_1) \mathcal{O}_{j_2}(x_2, u_2) \mathcal{O}_{j_3}(x_3, u_3)  \rangle = \\
&C_{j_1 j_2 j_3} \frac{(u_1 \cdot u_2)^{(j_1 + j_2 - j_3)/2} (u_1 \cdot u_3)^{(j_1 + j_3 - j_2)/2 } (u_2 \cdot u_3)^{(j_2 + j_3 - j_1)/2 }}{ |x_{12}|^{\Delta_{j_1} + \Delta_{j_2} - \Delta_{j_3} } |x_{13}|^{\Delta_{j_1} + \Delta_{j_3} - \Delta_{j_2} } |x_{23}|^{\Delta_{j_2} + \Delta_{j_3} - \Delta_{j_1} } }.
\end{split}
\end{equation}
In the following, we will calculate this overall constant when the two operators are heavy and the third one is light.
We will choose a configuration such that $j_1$ and $j_2$ are large with $j_3$ held fixed. To be specific, let us choose $j_1 = j + q, j_2 = j$ where $j \rightarrow \infty$ while $\hat{j} = j/ N$ and $q$ are held fixed. The correlation function may be explicitly computed using techniques similar to what we used to calculate the two-point function earlier 
\begin{equation}
\begin{split}
&\langle \mathcal{O}_{j_1}(x_1, u_1) \mathcal{O}_{j_2}(x_2, u_2) \mathcal{O}_{j_3}(x_3, u_3)  \rangle  \\
& =\frac{1}{Z}\int \cD \phi \cD \sigma \left( u_1 \cdot\phi (x_1) \right)^{j + q} \left( u_2 \cdot\phi (x_2) \right)^{j} \left( u_3 \cdot\phi(x_3) \right)^{j_3} e^{-\tfrac{C}{2} \int d^d y d^d z \tfrac{\phi^K(y) \phi^K (z)}{|y-z|^{d+s}} - \frac{1}{2} \int d^d x \sigma \phi^K \phi^K (x)} \\
&=  n_{j + q, j, j_3} \int  \cD \sigma \left(u_1 \cdot u_2 G_{12} \right)^{j + (q - j_3)/2} \left(u_1 \cdot u_3 G_{13} \right)^{(j_3 + q)/2} \left(u_2 \cdot u_3 G_{23} \right)^{(j_3 - q)/2} \\
& \hspace{2cm} \times e^{-\frac{N}{2} \log\det\lr{\frac{C}{|x-y|^{d+s}}+\sigma(x) \delta^d(x-y))}} 
\end{split}
\end{equation}
where in the last line, we just did the Wick contractions which gave rise to the combinatorial factor 
\begin{equation}
n_{j + q, j, j_3} = \frac{(j + q)! j! j_3!}{\left( j + \frac{q - j_3}{2} \right)! \left( \frac{q + j_3}{2} \right)! \left( \frac{j_3 - q}{2} \right)! }.
\end{equation}
Note that this three point function is only nonzero when $-j_3 \leq q \leq j_3$ and $j_3 + q$ is even. The Green's function $G_{ij} = G(x_i, x_j; \sigma)$ is the two point function in the presence of a non-trivial $\sigma$. However, notice that at large $j$ and $N$, the saddle point will only be affected by the factor of $(G_{12})^j$ in the prefactor above which gives the same exponent as in \eqref{TwoPointExact}. Therefore the large $N$, large $j$ saddle point is the same as before, and the three-point function is given by simply plugging in the previous saddle point solution into the above equation. Then, using \eqref{FullGreenLim} and \eqref{FunctDetRes}, we get 
\begin{equation}
\begin{split}
&\langle \mathcal{O}_{j_1}(x_1, u_1) \mathcal{O}_{j_2}(x_2, u_2) \mathcal{O}_{j_3}(x_3, u_3)  \rangle = n_{j + q, j, j_3} \left( \frac{\Gamma \left(\frac{d}{2} \right) \mu'(c_{\sigma})}{2 \pi^{d/2}} \right)^{j + \frac{j_3 + q}{2}} \delta^{2 N F(c_{\sigma}) + (2 j + q) \left(\mu - \left(\frac{d - s}{2} \right) \right)} \\
& \times \frac{\left(u_1 \cdot u_2 \right)^{j + (q - j_3)/2} \left(u_1 \cdot u_3 \right)^{(j_3 + q)/2} \left(u_2 \cdot u_3 \right)^{(j_3 - q)/2}}{|x_{12}|^{2 N F(c_{\sigma}) + \mu (2 j + q) - j_3 (\frac{d - s}{2} ) } |x_{13}|^{\mu q + j_3 (\frac{d - s}{2}) }  |x_{23}|^{-\mu q + j_3 (\frac{d - s}{2}) }}.
\end{split}
\end{equation} 
Note that this is scheme dependent through its dependence on $\delta$. To get a scheme independent result, we can choose  a normalization in which the coefficient of the two-point function is normalized to one. For that, we need to divide the above result by the square root of the coefficient of the two-point functions. For the large charge operators, the two-point function is normalized as
\begin{equation}
\langle \mathcal{O}_{j}(x_1, u_1) \mathcal{O}_{j}(x_2, u_2)\rangle = \left( \frac{\Gamma \left(\frac{d}{2} \right) \mu'(c_{\sigma})}{2 \pi^{d/2}} \right)^{j} \frac{\left(u_1 \cdot u_2 \right)^{j} j! \delta^{2 N F(c_{\sigma}) + j (2\mu - (d - s)) }}{|x_{12}|^{2 N F(c_{\sigma}) + 2 j \mu} }
\end{equation}
as we found earlier, while for the light operator, we have the usual normalization 
\begin{equation}
\langle \mathcal{O}_{j_3}(x_1, u_1) \mathcal{O}_{j_3}(x_2, u_2)\rangle = \frac{\left(u_1 \cdot u_2 \right)^{j_3} j_3! C_{\phi}^{j_3}}{|x_{12}|^{j_3 (\frac{d - s}{2})} }.
\end{equation}
Then the normalized coefficient of three-point function is 
\begin{equation}
a_{j+q, j, j_3} = \frac{n_{j + q, j, j_3}}{\sqrt{(j + q)! j! j_3!}} \left( \frac{\Gamma \left(\frac{d}{2} \right) \mu'(c_{\sigma})}{2 \pi^{d/2} C_{\phi}} \right)^{\frac{j_3}{2}} = \frac{\sqrt{j_3!}}{\left( \frac{q + j_3}{2} \right)! \left( \frac{j_3 - q}{2} \right)!} \left( \frac{2^{s - 1}\Gamma \left(\frac{d}{2} \right) \Gamma \left(\frac{s}{2} \right) \mu'(c_{\sigma}) N \hat{j}}{ \Gamma \left(\frac{d - s}{2} \right) }\right)^{\frac{j_3}{2}}
\end{equation}
where we already took the large $N$ limit. The $c_{\sigma}$ in the above expression is the one that solves the saddle point equation \eqref{SaddleEqRes}. At large $\hat{j}$, the OPE coefficient has the following scaling 
\begin{equation} \label{OPECoeffThreeP}
a_{j+q, j, j_3} =  \frac{\sqrt{j_3!}}{\left( \frac{q + j_3}{2} \right)! \left( \frac{j_3 - q}{2} \right)!} \left(-\frac{N \Gamma \left(\frac{s}{2} \right)   }{\Gamma \left(\frac{d - s}{2} \right)} \left(\frac{\pi}{ s \Gamma \left(\frac{d}{2}\right) \sin \left(\frac{\pi  d}{s}\right)} \right)^{\frac{2 s }{d + s}}  \right)^{\frac{j_3}{2}} \left( \frac{\hat{j} \Gamma \left(\frac{d + s}{2} \right) }{\Gamma \left(-\frac{s}{2} \right)} \right)^{\frac{j_3 (d - s)}{2 (d + s)}}  + \dots 
\end{equation}

Next, we look at the four-point function of two large charge and two finite charge operators. For simplicity of presentation, let us choose the large charge operators to be $\mathcal{Z}^j$ and $\bar{\mathcal{Z}}^j$ with $ \mathcal{Z}^j = (\phi^1 + i \phi^2)^j $. Let us further choose the finite charge operators to have charge one. Then there are two possible four point functions we can consider. The first one is \footnote{We are now going to not write the factors of renormalization scale $\delta$ here anymore, which as we explained in the context of three-point function, may be absorbed in the definition of the operators.}
\begin{equation}
\begin{split}
\langle \mathcal{Z}^j(x_1) \bar{\mathcal{Z}}^j(x_2)  \phi^a(x_3) \phi^b (x_4)\rangle &= \delta^{ab} j! \int  \cD \sigma \left(2 G_{12} \right)^{j} G_{34} e^{-\frac{N}{2} \log\det\lr{\frac{C}{|x-y|^{d+s}}+\sigma(x) \delta^d(x-y))}}   \\
&= \frac{\delta^{ab} j!}{|x_{12}|^{2 N F(c_{\sigma}) + 2 j \mu}} \left( \frac{\Gamma \left(\frac{d}{2} \right) \mu'(c_{\sigma})}{ \pi^{d/2}} \right)^{j} G(x_3, x_4; \sigma^*)
\end{split}
\end{equation}
where $a,b$ are not equal to $1$ or $2$. So this four-point function is just proportional to the Green's function we found in \eqref{GreensFunctionCrossRatio}. As expected, this has the form of a CFT four-point function 
\begin{equation}
\begin{split}
&\langle \mathcal{Z}^j(x_1) \bar{\mathcal{Z}}^j(x_2)  \phi^a(x_3) \phi^b (x_4)\rangle =  \frac{\delta^{ab} j!}{|x_{12}|^{2 N F(c_{\sigma}) + 2 j \mu} |x_{34}|^{d - s} } \left( \frac{\Gamma \left(\frac{d}{2} \right) \mu'(c_{\sigma})}{ \pi^{d/2}} \right)^{j} \frac{C_{\phi} \Gamma \left( \frac{d - 2}{2} \right)}{ \left( X Y \right)^{\frac{d - s}{4}}} \\
& \times \sum_{l = 0}^{\infty} \left( l + \frac{d- 2}{2} \right)  C_{l}^{ \left( \frac{d- 2}{2} \right)} \left(\frac{X + Y - 1}{2 \sqrt{X Y}} \right)  \int \frac{d u}{2 \pi} \left( \frac{Y}{X} \right)^{i u} \frac{Q_{l} (u)}{1 + C_{\phi} c_{\sigma} \pi^{d/2} Q_{l} (u)}
\end{split}
\end{equation} 
with cross-ratios 
\begin{equation}
X = \frac{|x_3 - x_1|^2 |x_4 - x_2|^2}{|x_1 - x_2|^2 |x_3 - x_4|^2}, \hspace{1cm} Y = \frac{|x_3 - x_2|^2 |x_4 - x_1|^2}{|x_1 - x_2|^2 |x_3 - x_4|^2}.
\end{equation}

The other four-point function we can consider in this simple setting is 
\begin{equation}
\begin{split}
&\langle \mathcal{Z}^j(x_1) \bar{\mathcal{Z}}^j(x_2)  \mathcal{Z}(x_3) \bar{\mathcal{Z}}(x_4) \rangle = \\
& j! 2^{j + 1} \int  \cD \sigma \left( G_{12}^{j} + j G_{12}^{j-1} G_{14} G_{23} \right)  e^{-\frac{N}{2} \log\det\lr{\frac{C}{|x-y|^{d+s}}+\sigma(x) \delta^d(x-y))}}.
\end{split}
\end{equation}
Because of the explicit factor of $j$ upfront, the second term dominates. Then using \eqref{FullGreenLim} and \eqref{FunctDetRes} as before, we get 
\begin{equation}
\begin{split}
\langle \mathcal{Z}^j(x_1) \bar{\mathcal{Z}}^j(x_2)  \mathcal{Z}(x_3) \bar{\mathcal{Z}}(x_4) \rangle = \left( \frac{\Gamma \left(\frac{d}{2} \right) \mu'(c_{\sigma})}{ \pi^{d/2}} \right)^{j + 1} \frac{(j + 1)!}{|x_{12}|^{2 N F(c_{\sigma}) + 2 j \mu} |x_{34}|^{d - s}} X^{\frac{2 \mu - (d - s)}{4}} Y^{\frac{-2 \mu - (d - s)}{4}}.
\end{split}
\end{equation}
Let us also normalize this four-point function by dividing it by the two-point function coefficients, so that we can extract the OPE coefficients in $13 \rightarrow 24$ channel  and compare it with what we got by calculating three-point functions 
\begin{equation}
\begin{split}
\langle \mathcal{Z}^j(x_1) \bar{\mathcal{Z}}^j(x_2)  \mathcal{Z}(x_3) \bar{\mathcal{Z}}(x_4) \rangle_{\textrm{norm.}} &= \left( \frac{\Gamma \left(\frac{d}{2} \right) \mu'(c_{\sigma})}{ 2 C_{\phi} \pi^{d/2}}  \right) \frac{ N \hat{j}}{|x_{12}|^{2 \Delta_j} |x_{34}|^{d - s}} X^{\frac{2 \mu - (d - s)}{4}} Y^{\frac{-2 \mu - (d - s)}{4}}  \\
&= \frac{1}{(x_{13} x_{24})^{\Delta_j + \Delta_{\phi}}} \left( \frac{x_{34}}{x_{12}} \right)^{\Delta_j - \Delta_{\phi}} \sum_{\Delta, s} a^2_{\Delta, s} X^{\frac{\Delta - s}{2}} g_{\Delta, s} (X, Y)
\end{split}
\end{equation}
where we expanded the four-point function into conformal blocks $g_{\Delta,s}$ in the $13 \rightarrow 24$ channel  (see for instance \cite{Dolan:2011dv}). Now that we have the four-point function, it is possible to extract all the OPE data from it, but here we will just calculate the OPE coefficient of the leading operator that appears in $13 \rightarrow 24$ channel. It should be a scalar operator with charge $j + 1$ and should correspond to the leading term in the four-point function in the limit $X \rightarrow 0$ and $Y \rightarrow 1$. The blocks are normalized such that $g_{\Delta,s} (X,Y) = 1 + \dots $ in the limit $X \rightarrow 0, Y \rightarrow 1$. Then the leading operator has the following dimension and OPE coefficient
\begin{equation}
\Delta_{j + 1} = \Delta_j + \mu \hspace{1cm} a^2_{j, 1, j + 1} = \left( \frac{ 2^{s - 1} \Gamma \left(\frac{d}{2} \right) \Gamma \left(\frac{s}{2} \right) \mu'(c_{\sigma})}{ \Gamma \left(\frac{d - s}{2} \right) }\right) N \hat{j}.
\end{equation}
This OPE coefficient agrees with what we found using the three-point function calculation \eqref{OPECoeffThreeP} for $j_3 = q = 1$. The result for the scaling dimension implies that at large $j$, the derivative of the dimension with respect to $j$ is $\mu$ 
\begin{equation}
\Delta_{j + 1} - \Delta_j = \frac{\partial \Delta_j }{\partial j}  = \mu. 
\end{equation}
This is in agreement with the structure of the result for the scaling dimension that we found earlier, as can be seen as follows 
\begin{equation}
\frac{\partial \Delta_j}{\partial j}  = \frac{\partial }{\partial j} \left( N F(c_{\sigma}) + \mu j \right) = \mu +  \left(N F'(c_{\sigma}) + \mu'(c_{\sigma}) j \right) \frac{\partial c_{\sigma} }{\partial j} = \mu
\end{equation} 
where we used the saddle point equation \eqref{SaddleEqRes}.

\subsection{Crossover to the short range regime}
\label{cross}
As we mentioned earlier, at a certain critical $s_*$, the behavior of the long range model is expected to cross over to that of the short-range $O(N)$ model. For recent discussions of this crossover see \cite{Behan:2017emf, Behan:2017dwr, Chai:2021arp}. 
Recall that $s_* = 2 - 2 \gamma_{\phi}^{\rm SR}$ where $\gamma_{\phi}^{\rm SR}$ is the anomalous dimension of $\phi$ at the short-range fixed point. However, in the large $N$ expansion, $\gamma_{\phi}$ is of order $1/N$, so in the regime we are working in, the crossover must happen at $s = 2$. In this subsection, we will study how the dimensions of the large charge operators behave near $s = 2$, and how the scaling dimension of the large charge operators may cross over from the long range to the short range behavior.  

\begin{figure} 
\centering
\begin{subfigure}{0.48\textwidth}
\includegraphics[width = \textwidth]{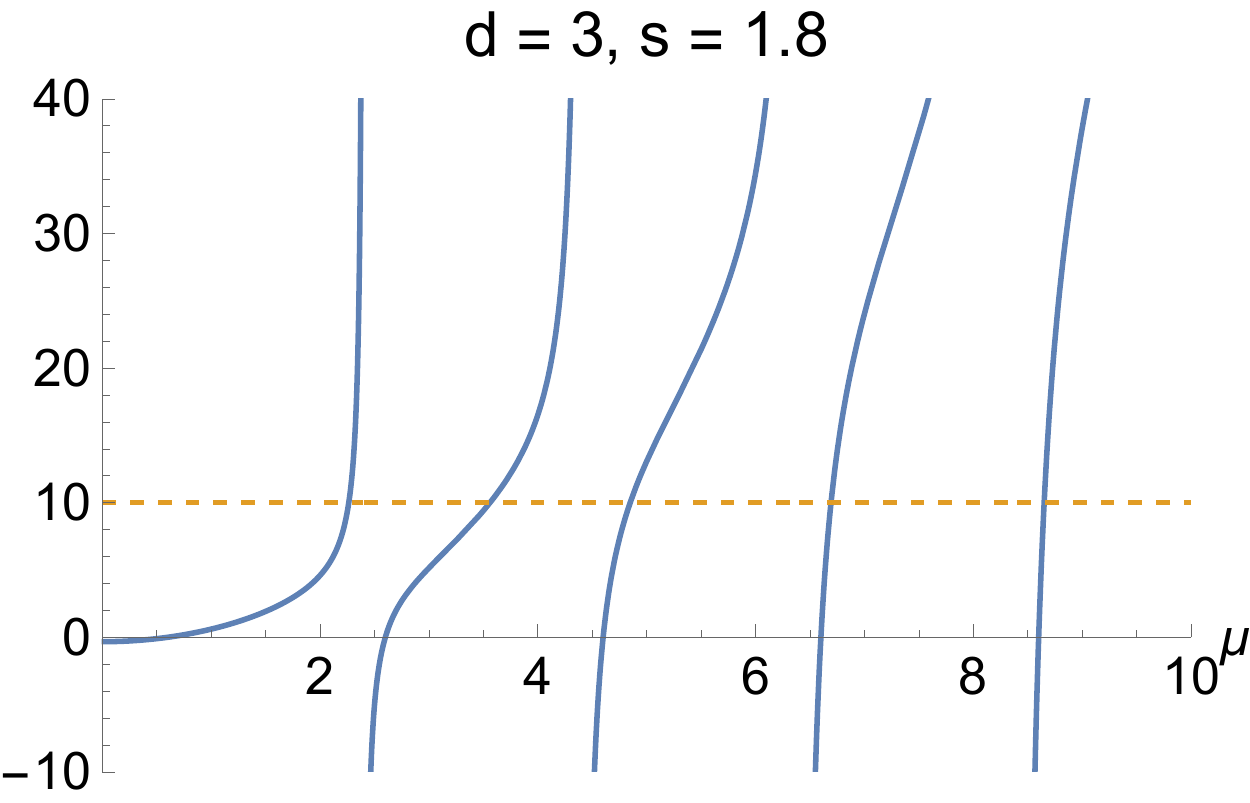}
\end{subfigure}
\begin{subfigure}{0.48\textwidth}
\includegraphics[width = \textwidth]{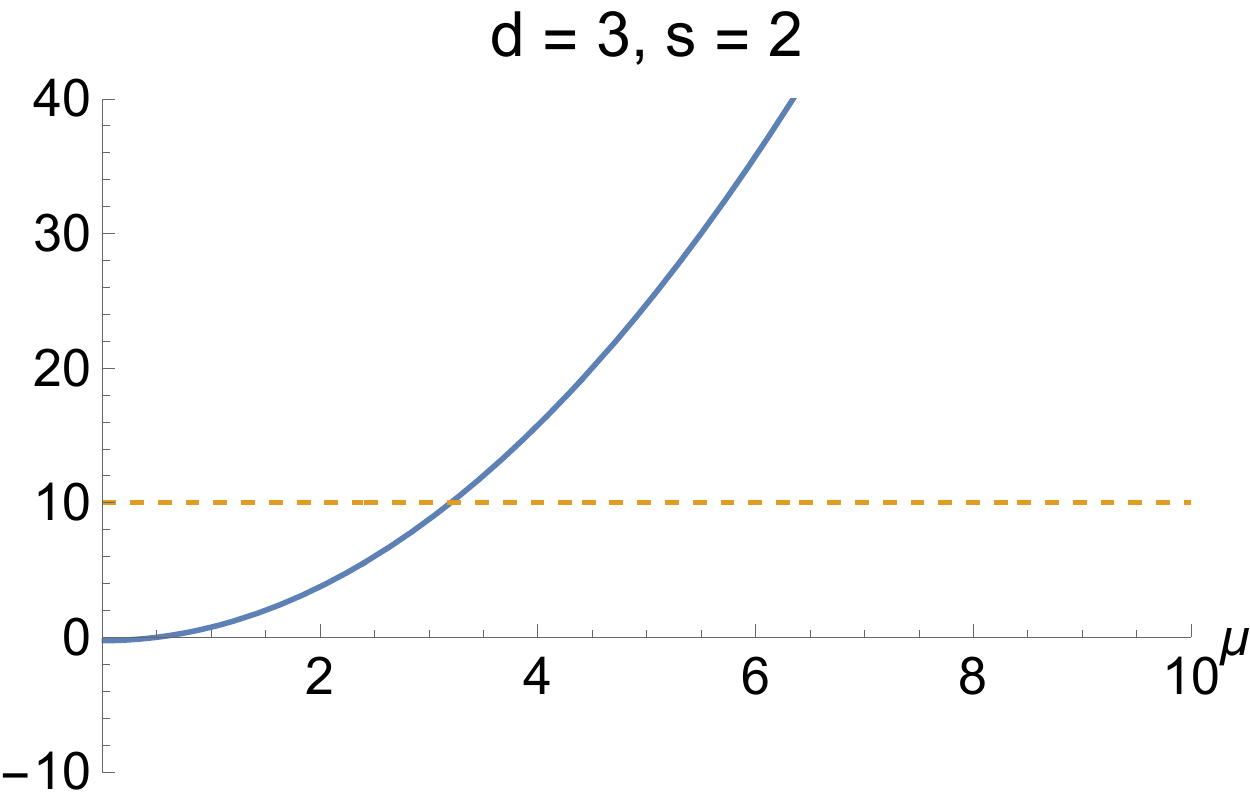}
\end{subfigure}
\caption{The plot of \eqref{EqWCSigma} for $d = 3$ at $s = 1.8$ and $s = 2$. The horizontal dashed line is the line $c_{\sigma} = 10$ and for this value of $c_{\sigma}$, the solution $\mu$ of \eqref{EqWCSigma} is given by the point where this dashed line intersects the curve. It is clear that there are several branches of solution for $s < 2$ while there is only one for $s = 2$.}
\label{figEq230}
\end{figure}

\begin{figure} 
\centering
\begin{subfigure}{0.48\textwidth}
\includegraphics[width = \textwidth]{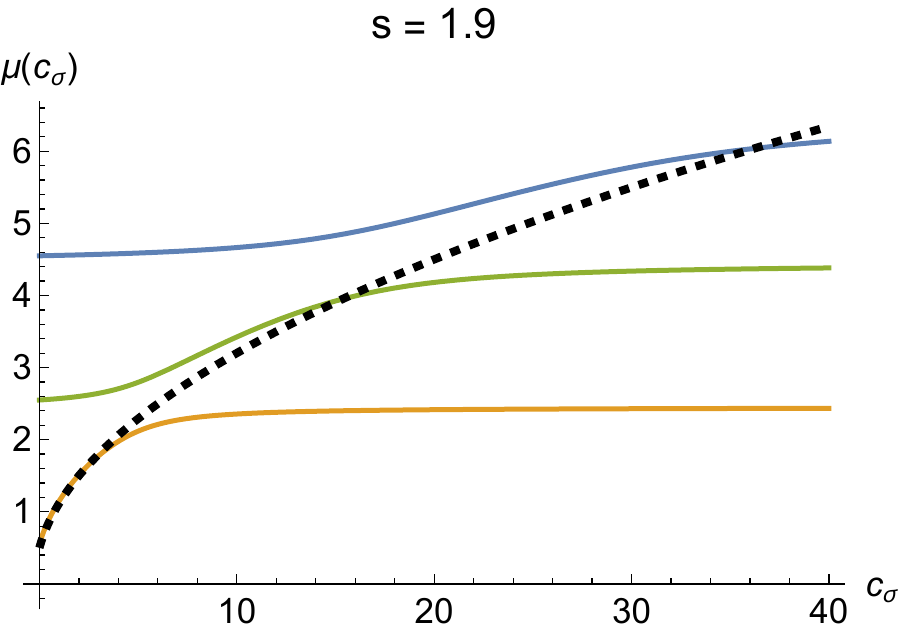}
\end{subfigure}
\begin{subfigure}{0.50\textwidth}
\includegraphics[width = \textwidth]{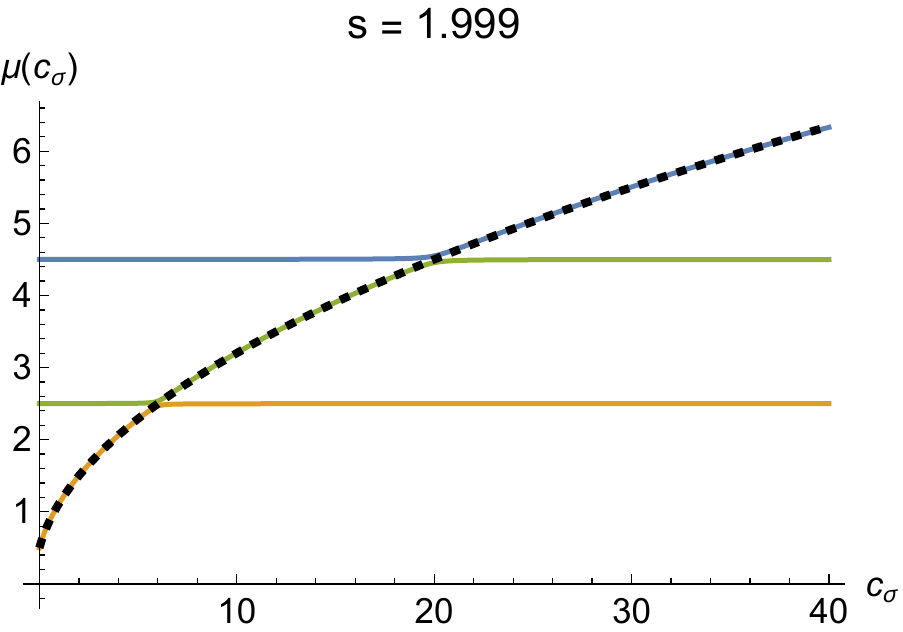}
\end{subfigure}
\caption{The numerical result for various branches of the solution $\mu(c_{\sigma})$ for $s = 1.9$ and for $s = 1.999$ in $d = 3$ dimensions. The dashed line is the $s = 2$ result (\ref{museq2}), so it is clear that $s = 2$ result arises by ``gluing" different branches. }
\label{mu3dsClose2}
\end{figure}

Let us start by observing that the solution of \eqref{EqWCSigma}, with $u=i\mu/2$ (where $\mu=\mu(c_{\sigma})$ will be physically related to the chemical potential on the cylinder), has several branches for any $s < 2$ , while for $s = 2$, it has a single solution given by 
\begin{equation} \label{museq2}
\mu(c_{\sigma}) = \sqrt{c_{\sigma} + \left(\frac{d}{2} - 1 \right)^2},
\end{equation}
see figure \ref{figEq230}. At small $c_{\sigma}$, the values of $\mu$ on the various branches go as $\mu(c_{\sigma})=\frac{d-s}{2}+2n+O(c_{\sigma})$  with $n=0,1,2,\ldots$. 
One can see that the small $c_{\sigma}$ expansion of (\ref{museq2}) matches what we get by just plugging in $s = 2$ in the small $c_{\sigma}$ expansion in  \eqref{RootLimits}, which gives the value of $\mu(c_{\sigma})$ on the first branch (the one with smallest $\mu$). But at large $c_{\sigma}$, the above $s=2$ result goes like $c_{\sigma}^{1/2}$ which is very different from the large $c_{\sigma}$ expansion in \eqref{RootLimits}. 
In particular, for any $s < 2$ the result \eqref{RootLimits} saturates below $(d + 2)/2$, while the $s = 2$ solution \eqref{museq2} crosses that point at $c_{\sigma} = 2 d$ and keeps growing. So if we always stay on the first branch, the function $\mu(c_{\sigma})$ can only have a smooth transition from $s < 2$ to the $s = 2$ behavior for $c_{\sigma} < 2 d$, and beyond this value one may expect that the higher roots of \eqref{EqWCSigma} should play a role. To get further intuition, it is useful to plot the solutions for various branches of $\mu$ as $s$ approaches 2 (see figure \ref{mu3dsClose2}). One can see that the $s = 2$ result (\ref{museq2}) arises essentially by ``gluing" portions of different branches as $s$ approaches $2$ from below. Note that the function $\mu(c_{\sigma})$ on the first branch, as we approach $s=2$, tends to develop a kink at $c_{\sigma} = 2 d$. This becomes a true kink at $s=2$, with $\mu(c_{\sigma})$ turning to a constant beyond that value. Similar kinks appear on the higher branches.\footnote{The position of the kinks at $s=2$ is given by $c_{\sigma}=2(d+2n)(n+1)$, $n=0,1,2,\ldots$.} 

\begin{figure} 
\centering
\begin{subfigure}{0.48\textwidth}
\includegraphics[width = \textwidth]{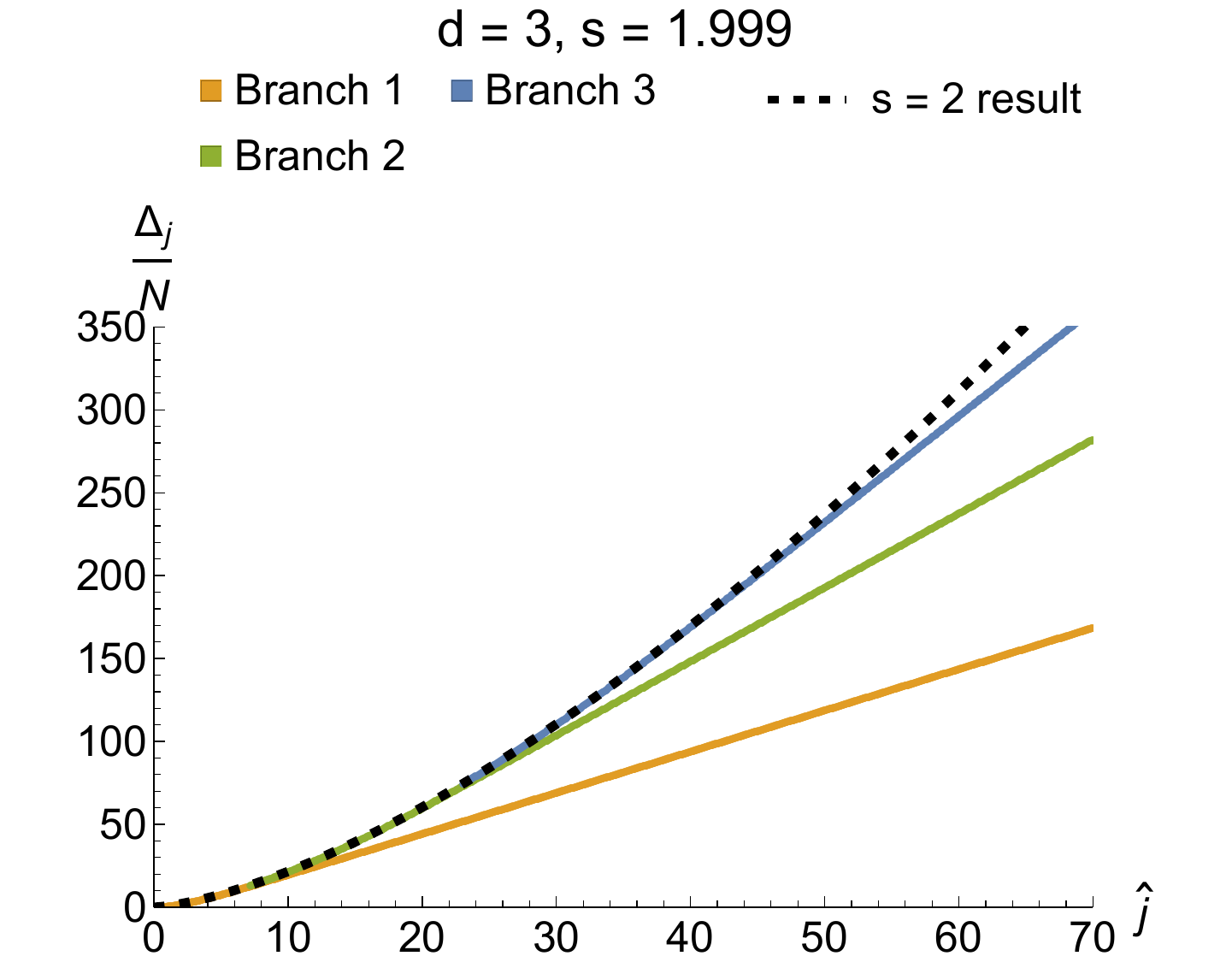}
\end{subfigure}
\begin{subfigure}{0.48\textwidth}
\includegraphics[width = \textwidth]{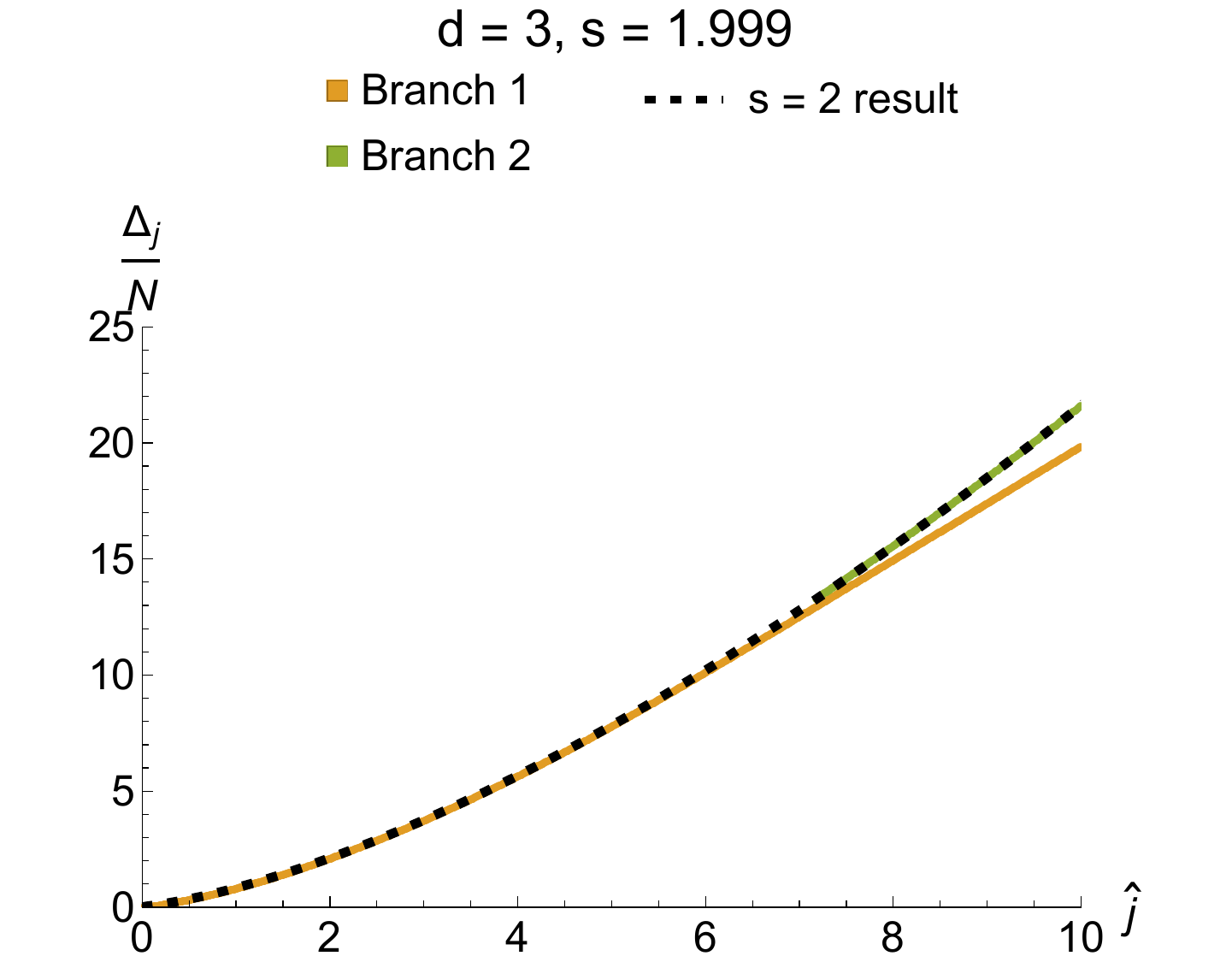}
\end{subfigure}
\caption{The numerical results for the dimension $\Delta_j$ obtained from various different branches of $\mu(c_{\sigma})$ for $d = 3, s = 1.999$. In the right plot, we just zoom in to the small $\hat{j}$ region. }
\label{FigureDeltawj1999}
\end{figure}

When we compute the scaling dimension by extremizing $\Delta_j =N(F(c_{\sigma})+\hat{j}\mu(c_{\sigma}))$ with respect to $c_{\sigma}$, we expect the higher branches of $\mu$ to lead to additional solutions to the saddle point equation. At small $\hat{j}$, it is easy to see that this leads to a tower of solutions with $\Delta_j^{(n)}/N=(\frac{d-s}{2}+2n)\hat{j}+O(\hat{j}^2)$. For finite $\hat{j}$, one can find these solutions numerically. Given the above discussion, we should see that the $s = 2$ behavior for $\Delta_j$ arises by ``gluing" the contributions of the saddles obtained from different branches.\footnote{Note that the functional determinant $F(c_{\sigma})$ has a smooth limit as $s \rightarrow 2$, which can for instance be seen by setting $s = 2$ in \eqref{FSmallcsigma} and \eqref{LargecF} and checking that it agrees with the results in \cite{Giombi:2020enj}.}  This is indeed what we find, as shown in figure \ref{FigureDeltawj1999}. The $s = 2$ case was considered in \cite{Alvarez-Gaume:2019biu, Giombi:2020enj}. In $d = 3$ and at large $\hat{j}$ the result behaves as 
\begin{equation} \label{s=2Largej}
\frac{\Delta_j}{N} = \frac{2}{3} \hat{j}^{\frac{3}{2}} + \frac{1}{6}  \hat{j}^{\frac{1}{2}} +  O \left(\frac{1}{\hat{j}^{\frac{1}{2}}} \right). 
\end{equation}
In figure \ref{FigureDeltawj1999}, we just plot the result in \eqref{s=2Largej},
since, as observed for instance in \cite{Giombi:2020enj}, the large $\hat{j}$ expansion gives a very good approximation to the true numerical value even down to relatively low $\hat{j}$. Note that, as shown in figure \ref{FigureDeltawj1999} to the right, the solution coming from the first branch smoothly goes to the $s = 2$ behavior for $\hat{j} < \hat{j}_{\textrm{crit.}}$. We can get an estimate for this critical value of $\hat{j}$. As we saw above, the $s = 2$ result for $\mu (c_{\sigma})$ starts diverging from the $s < 2$ result at $c_{\sigma} = 6$ in $d = 3$. When $s = 2$ and $d = 3$,  $c_{\sigma}$ is related to $\hat{j}$ as \cite{Alvarez-Gaume:2019biu, Giombi:2020enj}
\begin{equation}
c_{\sigma} = \hat{j} - \frac{1}{12} + O \left( \frac{1}{\hat{j}} \right).
\end{equation}
So we expect  the curve for the $s = 2$ result to diverge from the $s \lesssim 2$ result at around $\hat{j} \sim 6$. Beyond this value, the $s=2$ behavior is instead well approximated by the saddle obtained from the second branch of $\mu$, until we reach another critical value of $\hat{j}$ around $\hat{j}\sim 20$, and so on. Note that in the strict $s\rightarrow 2$ limit, each branch produces a solution to the saddle point equation only within a certain interval of $c_{\sigma}$ (and corresponding $\hat{j}$), outside of which $\mu$ becomes a constant (see figure \ref{mu3dsClose2}), which does not allow for solutions to $F'(c_{\sigma})+\hat{j}\mu'(c_{\sigma})=0$. Therefore, in the $s\rightarrow 2$ limit, the short range behavior $\Delta_j\sim \frac{2}{3}\hat{j}^{\frac{3}{2}}$ is indeed reproduced by ``gluing" the saddle point solutions obtained from the different branches.\footnote{It would be interesting to see if this merging of the branches can be interpreted as some kind of operator mixing. Indeed, it is natural to think of the solutions for $\Delta_j$ obtained from the higher branches as the dimensions of operators with the same charge but higher bare dimensions. For instance, on the second branch we have $\Delta_j/N = \left(\frac{d-s}{2}+2\right)\hat{j}+\ldots$ at small $\hat{j}$, which could be viewed as the dimension of an operator of the schematic form $\sim \left(\partial^2 (\phi^1+i \phi^2)\right)^j$. While at small $\hat{j}$ the scaling dimensions on different branches are well separated, at sufficiently large $\hat{j}$ and $s\rightarrow 2$ they can approach each other, and mixing may occur.} However, for $s<2$ and infinite $N$, the dominant behavior always comes from the first branch, which in particular gives scaling dimensions that go as $\Delta_j \sim \frac{d + s}{2} j$ at large $\hat{j}$. 

The picture we described above applies in the infinite $N$ limit we studied in this paper, where the transtion to the short range regime should happen at precisely $s=2$. If we include $1/N$ corrections, however, the role of the higher branches should become important slightly below $s=2$, since the crossover is expected to happen at $s^*=2-2\gamma_{\phi}^{SR}=2-O(1/N)<2$. It would be interesting to compute the subleading corrections to the scaling dimensions by including the determinant of the fluctuations around the saddle points, and further clarify how the transition to the short range regime works within the large charge sector.

\subsection{The long range model in $d=1$}
Let us now consider the special case of the $d = 1$ long range $O(N)$ model. Contrary to the usual short range case, there is a non-trivial fixed point for the one-dimensional long range model in the range $0 < s < 1$ \cite{dyson1969, PhysRevLett.37.1577, Aizenman1988DiscontinuityOT, Aizenman1988, 2015CMaPh.334..719A}. At $s = 1$, we expect a crossover to the short range fixed point, which is the trivial zero temperature fixed point where all correlation functions become constant. So we expect the scaling dimensions to go to zero as $s$ approaches $1$. We will show that this is the case for the large charge operators that we have been considering. 

Let us start with the Green's function \eqref{GreensFunctionCrossRatio}. In $d = 1$, the two cross-ratios are related to each other and there is only one cross-ratio, which we can take to be $\chi$ defined by 
\begin{equation}
X = \chi^2, \hspace{1cm} Y = (1 - \chi)^2. 
\end{equation}
The makes the argument of the Gegenbauer polynomials equal to $1$ and then using \eqref{Gegenbauer1}, the result for the Green's function is 
\begin{equation} \label{GreensFunction1dCross}
\begin{split}
&G(x,y, \sigma^*)  \\
& = \sum_{l = 0}^{\infty} \frac{C_{\phi} \Gamma \left( \frac{d - 2}{2} \right) (2 l + d - 2) \Gamma \left( d - 2 + l \right) }{ 2 |x -y|^{1 - s} \left( \chi |1 - \chi| \right)^{\frac{1 - s}{2}}  \Gamma \left( d - 2 \right) l! }   \int \frac{d u}{2 \pi} \left( \frac{ \chi^2}{(1 - \chi)^2} \right)^{-i u} \frac{Q_{l} (u)}{1 + C_{\phi} c_{
\sigma} \pi^{d/2} Q_{l} (u)}\\
&= \sum_{l = 0,1} \frac{  \Gamma \left( \frac{1 - s}{2} \right)}{ 2^s  \Gamma \left( \frac{s}{2} \right)  |x -y|^{1 - s} \left( \chi |1 - \chi| \right)^{\frac{1 - s}{2}}  }   \int \frac{d u}{2 \pi} \left( \frac{ \chi^2}{(1 - \chi)^2} \right)^{-i u}  \frac{Q_{l} (u)}{1 + C_{\phi} c_{
\sigma} \pi^{d/2} Q_{l} (u) }.
\end{split}
\end{equation}
In $d = 1$, the prefactor of the integral vanishes unless $l = 0$ or $1$, so the sum collapses to only those two terms. In the next section, we will show that all these calculations may also be done by mapping to a cylinder, $R \times S^{d - 1}$. Then the sum over $l$ comes from summing over angular momentum modes on the sphere. However in $d = 1$, there is no sphere, so the sum over $l$ must collapse. 

Similarly for the functional determinant \eqref{FunctDetRes}, we have 
\begin{equation} \label{Fc1d}
F(c_{\sigma}) = \sum_{l = 0,1} \int_{-\infty}^{\infty} \frac{d u}{2 \pi}  \log \left( 1 + C_{\phi} c_{\sigma} \pi^{d/2} Q_{l} (u) \right) .
\end{equation}
At large $c_{\sigma}$, this goes like \eqref{FcLargec1dApp}  
\begin{equation} \label{LargecOneD}
F(c_{\sigma}) = \frac{c_{\sigma}^{1/s}}{\sin \left( \frac{\pi}{s} \right)} \left( 1 - \frac{1 - s^2}{24 c_{\sigma}^{2/s} }  + ... \right).
\end{equation}
Since we don't have an infinite sum over $l$, it is also possible to extract the large $c_{\sigma}$ behavior directly from \eqref{Fc1d}. To do that, we first differentiate $F(c_{\sigma})$ with $c_{\sigma}$ and expand at large $u$. We then rescale the integration variable $u \rightarrow c_{\sigma}^{1/s} \ u$ and then expand at large $c_{\sigma}$. Finally we perform the integral over $u$ and then the integrate back over $c_{\sigma}$. The two results of course agree. At generic values of $c_{\sigma}$, it is possible to evaluate $F(c_{\sigma})$ numerically using \eqref{FcReg}, with sum now only running over $l = 0$ and $l = 1$. 

\begin{figure} 
\centering
\begin{subfigure}{0.51\textwidth}
\includegraphics[width = \textwidth]{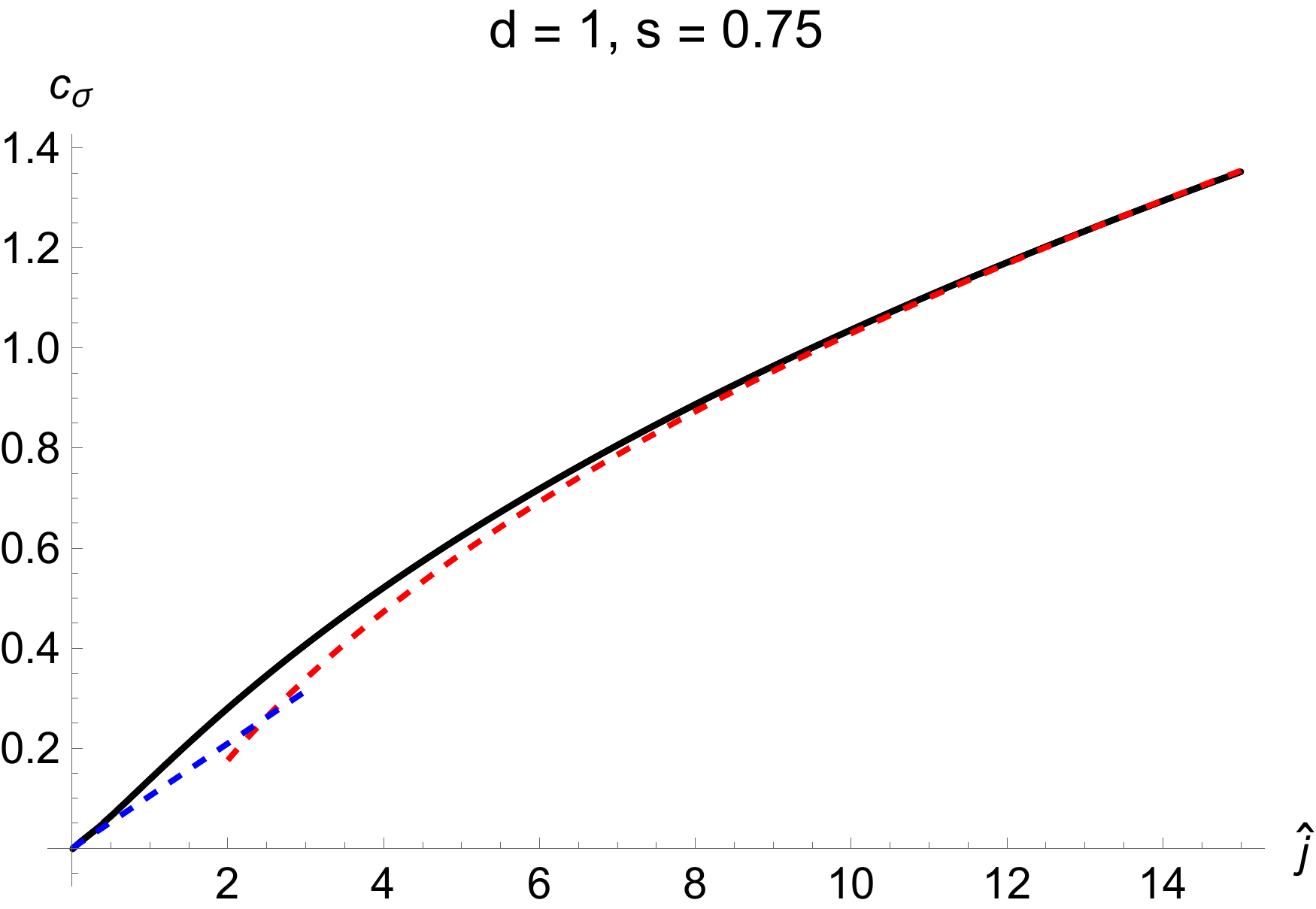}
\end{subfigure}
\begin{subfigure}{0.48\textwidth}
\includegraphics[width = \textwidth]{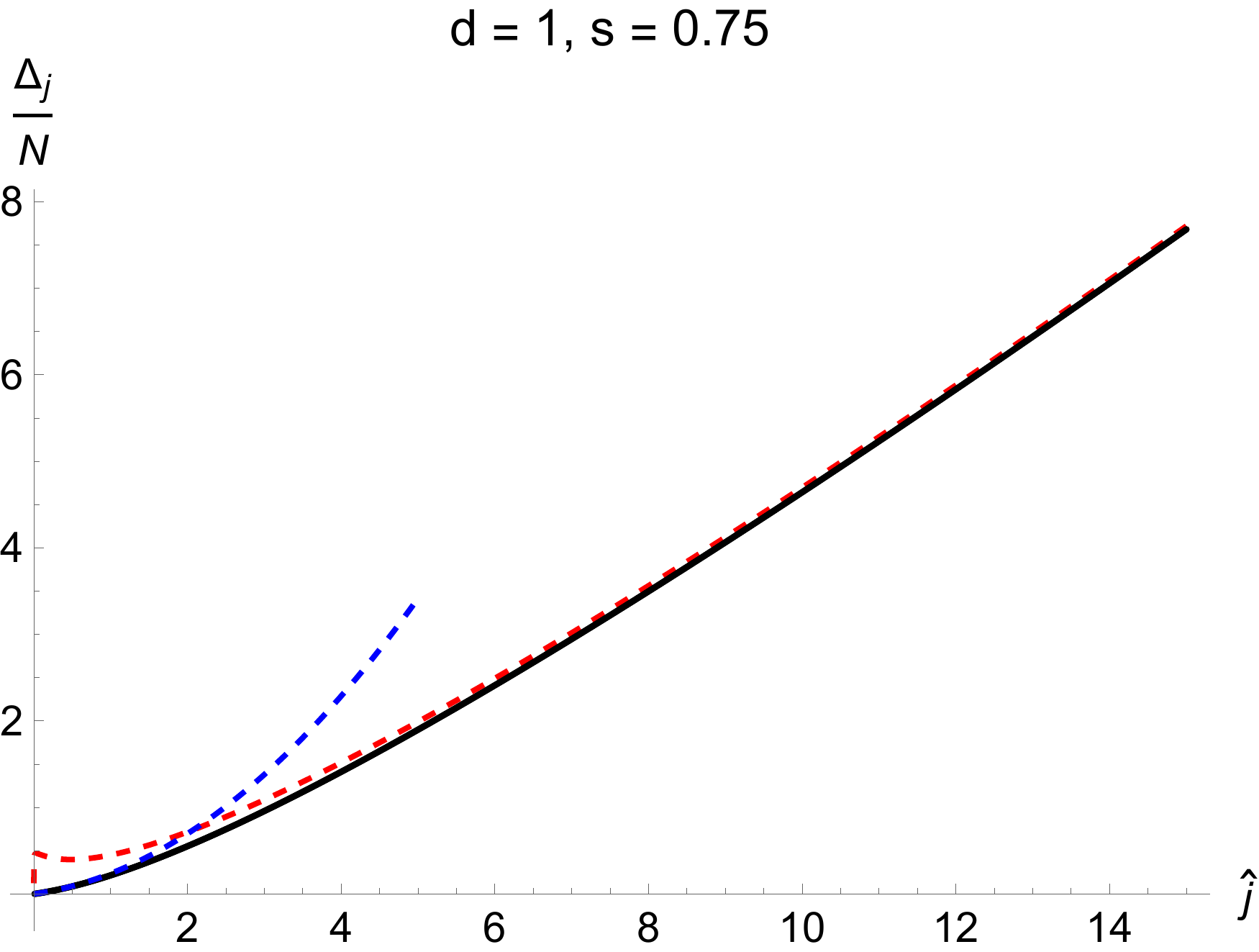}
\end{subfigure}
\caption{The numerical results for the dimension $\Delta_j$ and the solution to the saddle point equation for $d = 1, s = 0.75$. In both plots, black line represents the numerical results, the dashed red line is the analytical result in a large $\hat{j}$ expansion and the dashed blue line is the analytical result in a small $\hat{j}$ expansion. }
\label{FigureDeltawj075}
\end{figure}

Combining \eqref{LargecOneD} with the large $c_{\sigma}$ behavior of $\mu(c_{\sigma})$ from \eqref{RootLimits}, we can solve the saddle point equation at large $c_{\sigma}$ 
\begin{equation} \label{LargeJCOneD}
c_{\sigma} = \left(\frac{j 2^{s+1} s \Gamma \left(\frac{s+1}{2}\right)}{\sqrt{\pi } \csc \left(\frac{\pi }{s}\right) \Gamma \left(-\frac{s}{2}\right)}\right)^{\frac{s}{s+1}} -\frac{2 \sin \left(\frac{\pi  s}{2}\right) \Gamma (s+1) \left(\pi  s \cot \left(\frac{\pi  s}{2}\right)+2 s (\psi ^{(0)}(s)+\gamma )+2\right)}{\pi  (s+1)} .
\end{equation}
Corrections to the above are of order $O(j^{- \frac{s}{1 + s}})$ at large $\hat{j}$. We can then use this result to get the dimensions of the large charge operators in a large $\hat{j}$ expansion 
\begin{equation}  \label{LargeJDeltaOneD}
\begin{split}
\frac{\Delta_j }{N} &= \frac{1 + s}{2} \hat{j} + \frac{2  (1 + s)}{ \sin \left( \frac{\pi }{s} \right)  s } \left( \frac{ \Gamma \left( \frac{1 + s}{2} \right)  s \sin \left( \frac{\pi}{s} \right)  \hat{j} }{ \Gamma \left(- \frac{s}{2} \right) \sqrt{\pi} } \right) ^{\frac{1}{1 + s}}  \\ 
&- \frac{ 2^{1 - s} \Gamma(s) \sin \left( \frac{\pi s}{2} \right)  \left(\pi  s \cot \left(\frac{\pi  s}{2}\right)+2 s (\psi ^{(0)}(s)+\gamma )+2\right)}{ s \pi \sin \left( \frac{\pi}{s} \right) } \left( \frac{ \Gamma \left( \frac{1 + s}{2} \right)  s \sin \left( \frac{\pi}{s} \right)  \hat{j} }{ \Gamma \left(- \frac{s}{2} \right) \sqrt{\pi} } \right) ^{\frac{1 - s}{1 + s}}.
\end{split}
\end{equation}
It is also possible to numerically solve the saddle point equation and hence find the dimension of the large charge operators. We plot the result for the saddle solution and the dimensions for $s = 0.75$ in figure \ref{FigureDeltawj075}. Note that the analytical large $\hat{j}$ results in \eqref{LargeJCOneD} and \eqref{LargeJDeltaOneD} work remarkably well.   

Now let us discuss the behavior of the model as $s$ approaches $1$. Let us start by recalling the small $c_{\sigma}$ results for $F(c_{\sigma})$ and for $\mu(c_{\sigma})$ when $s$ is close to $1$
\begin{equation}
\begin{split}
\mu(c_{\sigma}) &= \frac{1 - s}{2}  + \frac{2 c_{\sigma} }{\pi  (1 - s)} - \frac{4 c_{\sigma}^2}{\pi ^2 (1 - s)^3} + O(c_{\sigma}^3). \\
F(c_{\sigma}) &= -\frac{2 c_{\sigma}^2 }{\pi^2 (1 - s)^3}
\end{split}
\end{equation}
where we only wrote the leading order in $1-s$ result at each order in $c_{\sigma}$. Note that the expansion of $\mu(c_{\sigma})$ clearly breaks down unless $c_{\sigma} \ll (1-s)^2$. Assuming the expansion of $F(c_{\sigma})$ has a similar validity, we can solve the saddle point equation using these two expressions and we get
\begin{equation} \label{SmalljSaddSolOned}
\begin{split}
c_{\sigma} = \frac{\hat{j} (1 - s)^2 \pi}{2}, \hspace{1cm} 
\frac{\Delta_j }{N}  =  \frac{1 - s}{2} \hat{j} + \frac{1-s }{2} \hat{j}^2  + O(\hat{j}^3)
\end{split}
\end{equation}
So this is only really valid for $\hat{j} \ll 1$. But clearly in this regime, both $c_{\sigma}$ and $\Delta_j$ go to zero as $s \rightarrow 1$. This is consistent with the expectation that the dimensions should go to zero as $s$ approaches $1$. Next let us look at the large $c_{\sigma}$ expansions  
\begin{equation}
\begin{split}
\mu(c_{\sigma}) &= 1  - \frac{2 }{\pi   c_{\sigma}} + \frac{4 }{\pi ^2 c_{\sigma}^2} + O(c_{\sigma}^3). \\
F(c_{\sigma}) &= \frac{c_{\sigma}}{\pi (s-1)}.
\end{split}
\end{equation}
Clearly, $F'(c_{\sigma})$ diverges as $s \rightarrow 1$, while $\mu'(c_{\sigma})$ is finite. So there is no solution to the saddle point equation with large $c_{\sigma}$ close enough to $s = 1$. 

\begin{figure} 
\centering
\begin{subfigure}{0.50\textwidth}
\includegraphics[width = \textwidth]{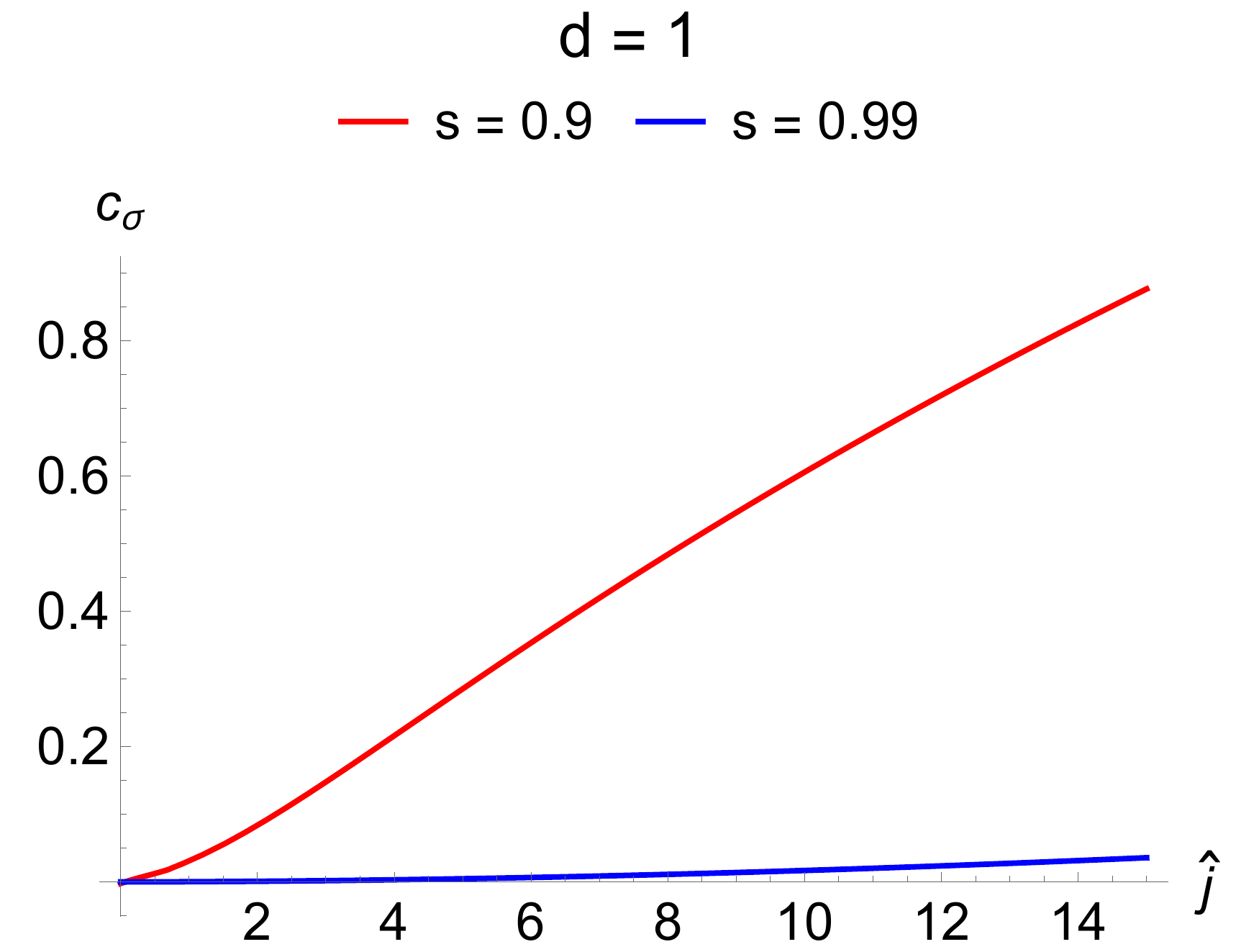}
\end{subfigure}
\begin{subfigure}{0.48\textwidth}
\includegraphics[width = \textwidth]{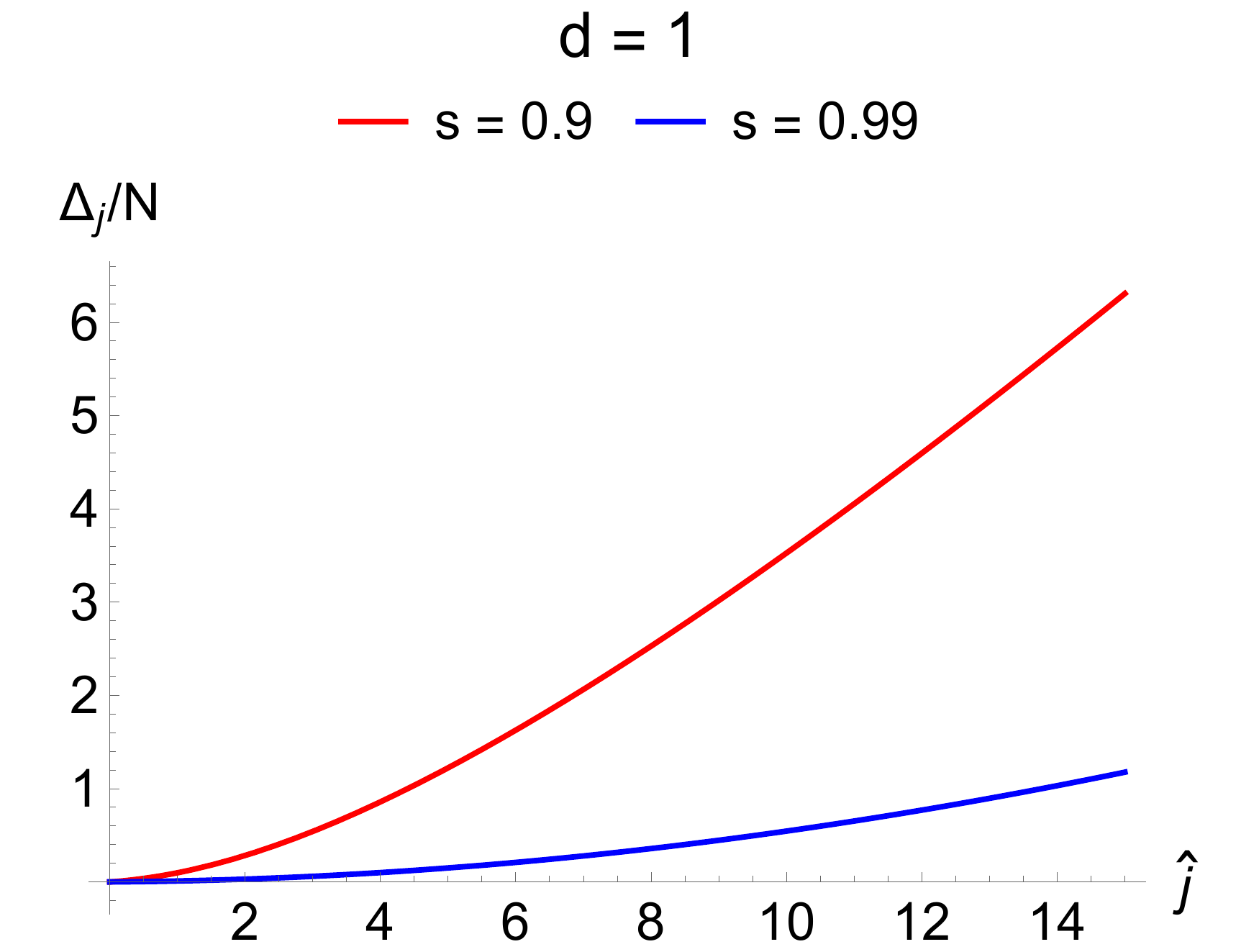}
\end{subfigure}
\caption{The numerical results for the dimension $\Delta_j$ and the solution to the saddle point equation in $d = 1$ for  $s = 0.9$ (red) and $s = 0.99$ (blue). As one can see, both $c_{\sigma}$ and $\Delta_j$ appear to approach zero as $s$ gets closer to $1$.} 
\label{FigureDeltawj099}
\end{figure}

\begin{figure} 
\centering
\begin{subfigure}{0.47\textwidth}
\includegraphics[width = \textwidth]{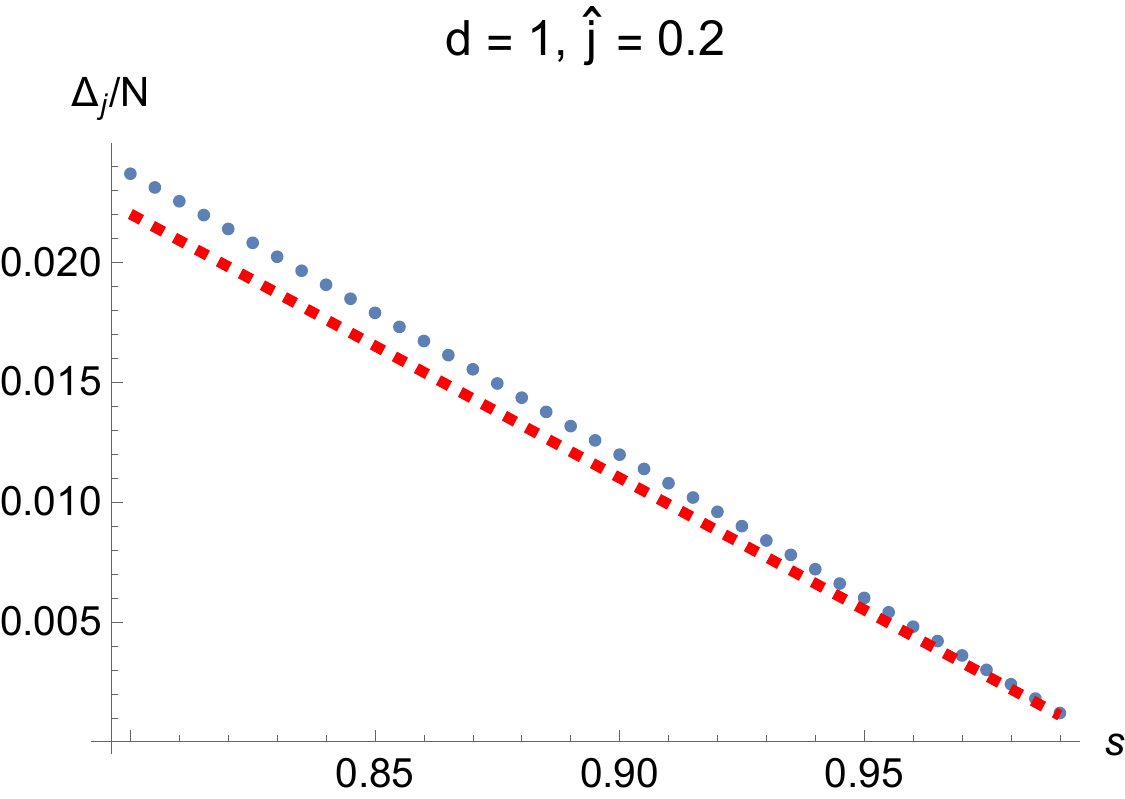}
\end{subfigure}
\begin{subfigure}{0.47\textwidth}
\includegraphics[width = \textwidth]{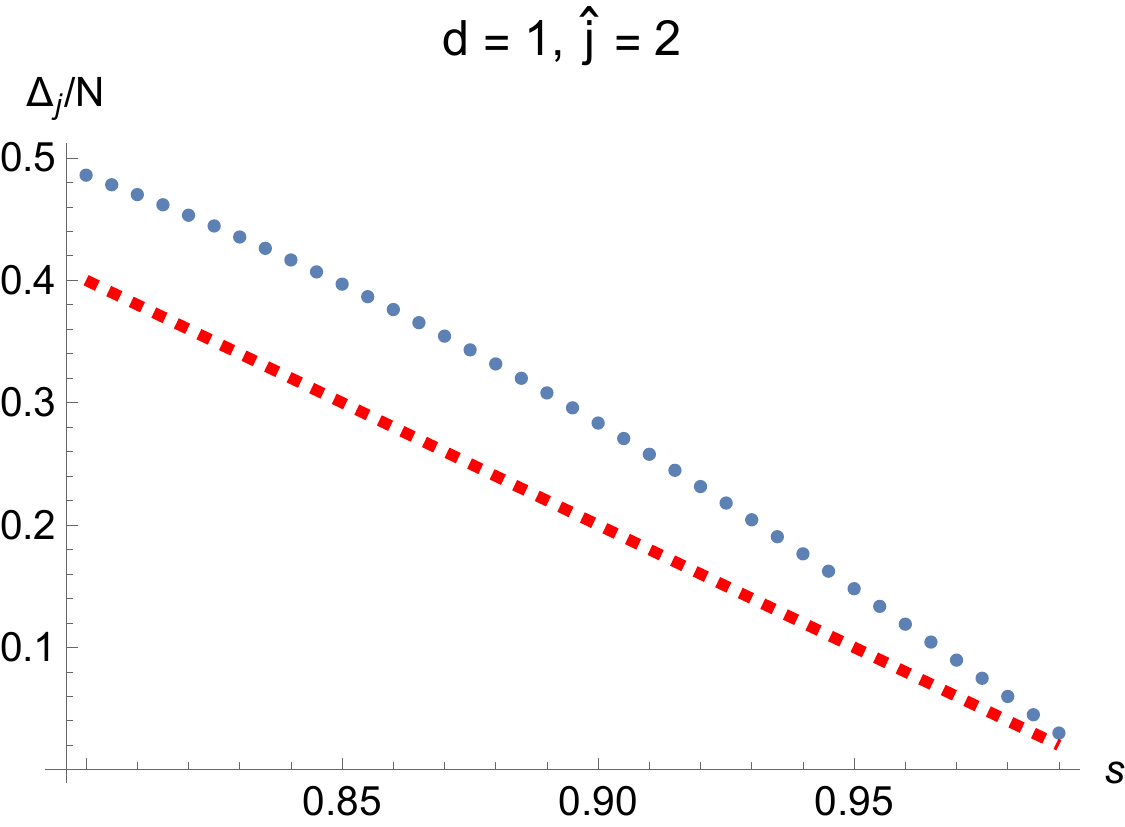}
\end{subfigure}
\caption{The numerical results for the dimension $\Delta_j$  as a function of $s$ for $\hat{j} = 0.2$ and for $\hat{j} = 2$. The blue dots are the numerical points while the red dashed line is the analytical result for small $\hat{j}$ \eqref{SmalljSaddSolOned}.} 
\label{FigureDeltaws}
\end{figure}

In order to clarify what happens for finite $\hat{j}$, let us then look at the numerics as $s$ gets closer to 1. In figure \ref{FigureDeltawj099}, we plot the solution to the saddle point equation for $c_{\sigma}$ and scaling dimension for $s = 0.9$ and $s = 0.99$, and it seems clear that as $s$ approaches $1$, both of these quantities approach zero. To see how they approach $0$, we can numerically evaluate $\Delta_j$ as a function of $s$ for a fixed $\hat{j}$. We show these results in figure \ref{FigureDeltaws}. The results seem to suggest the dimension goes to zero linearly as $1-s$ even when $\hat{j}$ is not too small. It would be interesting to clarify this further, perhaps using the non-local non-linear sigma model considered in \cite{Giombi:2019enr}, which has a perturbative fixed point in $d=1$ and $s=1-\epsilon$. 

\section{Scaling dimensions from the cylinder} \label{Sec:Cylinder}
In this section, we show that the scaling dimensions we calculated above can also be derived by studying the theory on a cylinder, 
which may be obtained by a Weyl transformation from the flat space. We will use this approach mainly as a check of the results we obtained above, so we will be brief. Such an approach has been used in several recent works  \cite{Hellerman:2015nra, Alvarez-Gaume:2016vff, Monin:2016jmo, Badel:2019oxl, Alvarez-Gaume:2019biu, Cuomo:2021cnb}. We will follow and generalize the approach used in \cite{Cuomo:2021cnb}, which studied large charge operators in a boundary conformal field theory. 

\subsection{$\epsilon$ expansion}
We start by considering the long range $O(N)$ model \eqref{ActionLongRangeQuadratic} in the vicinity of the lower critical value of $s$. For $s = \frac{d + \epsilon}{2}$, the model has a perturbative fixed point where the coupling is given by \cite{PhysRevLett.29.917}
\begin{equation}
g_* = \frac{(4 \pi)^{\frac{d}{2}} \Gamma \left(\frac{d}{2} \right)}{2 (N + 8)} \epsilon.
\end{equation}
We will work at this fixed point to leading order in $\epsilon$. It is convenient to think of the model as coming from the following model in $D = d + 2 - s$ dimensions, with interactions localized to the $d$ dimensional subspace \cite{Paulos:2015jfa}
\begin{equation}
S =  \frac{\Gamma \left(\frac{s}{2} \right)}{(4 \pi)^{1 - \frac{s}{2}}} \int d^{d} x \ d^{2 - s} w \frac{1}{2}  (\partial_{\mu} \Phi^I)^2 + \frac{g}{4} \int d^d x (\phi^I \phi^I)^2.
\end{equation} 
The field $\Phi^I$ is the $D$-dimensional extension of $\phi^I$ 
\begin{equation}
\Phi^I (x, w = 0) = \phi^I(x)
\end{equation}
and we restrict for simplicity to field configurations which depend on the extra coordinates only through $|w|$. We are interested in calculating the dimensions of fixed charge operators with a large charge $j$ while holding $\epsilon j$ fixed. For that purpose, we perform a Weyl transformation to the cylinder $R \times S^{D - 1}$. By the state-operator correspondence, the large charge operators on the plane are mapped to large charge states on the cylinder. We pick the following coordinates on $S^{D - 1}$
\begin{equation}
ds^2 = d \theta^2 + \sin^2 \theta d \Omega_{d - 1} + \cos^2 \theta d \Omega_{1 - s} 
\end{equation}
where $0 \le \theta \le \pi/2$. The limit $\theta \rightarrow \pi/2$ then brings us to the $d$ dimensional subspace our original model was defined on. We want to consider fixed large charge states on the cylinder. By ensemble equivalence, this can be done by introducing a fixed chemical potential $\mu$. Without loss of generality, we introduce this chemical potential for the $U(1)$ subgroup that rotates $\Phi^1$ and $\Phi^2$. In practice, the chemical potential may be implemented by having a background gauge field in the time direction \footnote{This is done by modifying the kinetic term in the action so that $\partial_0 \Phi^1 \rightarrow D_{0} \Phi^1 = \partial_0 \Phi^1 + i \mu \Phi^2$ and $\partial_0 \Phi^2 \rightarrow D_{0} \Phi^2 = \partial_0 \Phi^2 - i \mu \Phi^1$. } 
(see for instance \cite{Alvarez-Gaume:2019biu} for a related discussion) 
\begin{equation}
\begin{split}
S \rightarrow S + \frac{\Gamma \left(\frac{s}{2} \right)}{(4 \pi)^{1 - \frac{s}{2}}} \int_{R \times S^{D - 1}}  \left[ i \mu \left( \dot{\Phi}^1 \Phi^2 -  \dot{\Phi}^2 \Phi^1 \right) - \frac{\mu^2}{2} \left( (\Phi^1)^2 + (\Phi^2)^2 \right) \right]
\end{split}
\end{equation}
where dot represents the time derivative. We expand the field around the following ansatz \footnote{An equivalent way to introduce the chemical potential is to have a time dependent ansatz given by $\Phi^1 + i \Phi^2 = \sqrt{2}  f(\theta) e^{- i \mu t }$ where $t$ now is the Lorentzian time. } 
\begin{equation} \label{ClassicalAnsatz}
\Phi^1 + i \Phi^2 = \sqrt{2}  f(\theta), \hspace{1cm} \Phi^3 = \Phi^4 = \dots = \Phi^N  = 0.
\end{equation}
In this background, the classical action is 
\begin{equation}
\begin{split}
S_{\textrm{cl}} = & \frac{T 4^{\frac{s}{2}} \pi^{\frac{d}{2}} \Gamma \left(\frac{s}{2} \right) }{\Gamma \left(\frac{d}{2} \right) \Gamma \left(1 - \frac{s}{2} \right)} \int_{0}^{\pi/2} d \theta \left(\sin \theta \right)^{d - 1} \left(\cos \theta \right)^{1 - s} \left( (\partial_{\theta} f (\theta))^2   + \left( - \mu^2 + \frac{(d - s)^2}{4} \right) f (\theta)^2 \right) \\
&+ \frac{ 2 g T \pi^{\frac{d}{2}} }{\Gamma \left(\frac{d}{2} \right) } f (\theta)^4 \bigg|_{\theta = \frac{\pi}{2}}
\end{split}
\end{equation}
where $(d-s)^2/4$ comes from the conformal coupling on the cylinder and $T$ is the length of the cylinder along the Euclidean time direction, which is formally infinite. The variational principle gives the following equation of motion
\begin{equation} \label{ClassicalEOM}
\frac{1}{(\sin \theta)^{d - 1} (\cos \theta)^{1 - s} } \partial_{\theta} \left( (\sin \theta)^{d - 1} (\cos \theta)^{1 - s}  \partial_{\theta} f(\theta) \right) + \left( \mu^2 - \frac{(d - s)^2}{4} \right) f(\theta) = 0  
\end{equation}
along with the boundary condition
\begin{equation}
\left[ \frac{ 4^{\frac{s}{2}} \Gamma \left(\frac{s}{2} \right)  (\sin \theta)^{d - 1} (\cos \theta)^{1 - s}}{\Gamma \left(1 - \frac{s}{2} \right)}  \partial_{\theta} f(\theta) +  4 g f^3(\theta) \right]_{\theta = \frac{\pi}{2}} = 0 
\end{equation}
The solution that is regular at $\theta = 0$ is given by
\begin{equation} \label{ClassicalEOMSoln}
f(\theta) = v (\cos \theta)^{-\frac{d-s - 2 \mu}{2}} {}_2F_1 \left( \frac{d-s - 2 \mu}{4}, \frac{d + s - 2 \mu}{4}; \frac{d}{2}; - \tan^2 \theta \right).
\end{equation}
Using the boundary condition fixes (we can set $s = d/2$ for this calculation, to leading order in $\epsilon$) 
\begin{equation}
v^2 = - \frac{2^{\frac{d}{2} -1 }  \Gamma \left(\frac{3 d}{8}-\frac{\mu }{2}\right)^3 \Gamma \left(\frac{3 d}{8}+\frac{\mu }{2}\right)^3}{ g \Gamma \left(\frac{d}{4}\right)^2 \Gamma \left(\frac{d}{2}\right)^2 \Gamma \left(\frac{1}{8} (d-4 \mu )\right) \Gamma \left(\frac{1}{8} (d+4 \mu )\right)}.
\end{equation}

The scaling dimensions of the operators on the plane are then related to the energy on the cylinder and may be calculated by extremizing the following expression (this is essentially a Legendre transform from the free energy at fixed chemical potential to the free energy at fixed charge)
\begin{equation} \label{DimensionCylinder}
\begin{split}
\Delta_j &= \left[ \frac{S_{\textrm{cl}}}{T} + \mu j \right]_{\mu = \mu^*} = \left[ - \frac{ 2 g \pi^{\frac{d}{2}} f^4(\theta = \pi/2)}{ \Gamma \left(\frac{d}{2} \right) } + \mu j \right]_{\mu = \mu^*} \\
&= \left[  -\frac{2^{d-1} \pi^{\frac{d}{2}} \Gamma \left(\frac{3 d}{8}-\frac{\mu }{2}\right)^2 \Gamma \left(\frac{3 d}{8}+\frac{\mu }{2}\right)^2}{g \Gamma \left(\frac{d}{2} \right) \Gamma \left(\frac{1}{8} (d-4 \mu )\right)^2 \Gamma \left(\frac{1}{8} (d+4 \mu )\right)^2} + j \mu \right]_{\mu = \mu^*}
\end{split}
\end{equation} 
where we used the boundary condition and $\mu^*$ is the value of $\mu$ that extremizes the above expression
\begin{equation}
\frac{2^{d-1} \pi ^{d/2} \Gamma \left(\frac{3 d}{8}-\frac{\mu^* }{2}\right)^2 \Gamma \left(\frac{3 d}{8}+\frac{\mu^*}{2}\right)^2 \left[\psi ^{(0)}\left(\frac{3 d}{8}+\frac{\mu^*}{2}\right) -\psi ^{(0)}\left(\frac{1}{8} (d+4 \mu^* )\right) - \mu^* \rightarrow -\mu^* \right]}{\Gamma \left(\frac{d}{2}\right) \Gamma \left(\frac{1}{8} (d-4 \mu^* )\right)^2 \Gamma \left(\frac{1}{8} (d+4 \mu^* )\right)^2} = g j
\end{equation}
It is hard to find this extremal value analytically in general, but we can make progress in the limit of small and large $g j$. For small $g j$ we get 
\begin{equation} \label{DimEpsExpSmall}
\mu^* = \frac{d}{4} + \frac{g j}{2^{d - 2} \pi^{\frac{d}{2}} \Gamma \left(\frac{d}{2}\right) } + O(g j)^2 \implies  \Delta_j =  j \left( \frac{d}{4} + \frac{j \epsilon }{N + 8} + O(j \epsilon)^2 \right).
\end{equation}
For large $g j$, we get instead
\begin{equation} \label{DimEpsExpLarge}
\begin{split}
\mu^* &= \frac{3 d}{4} - \frac{2^{\frac{d+2}{3}} \pi ^{\frac{d-4}{6}} \left( \sin \left(\frac{\pi  d}{4}\right) \right)^{\frac{2}{3}} \left(\Gamma \left(\frac{d}{4}+1\right) \Gamma \left(\frac{3 d}{4}\right)\right)^{2/3}}{ (g j )^{\frac{1}{3}} \Gamma \left(\frac{d}{2}\right)} \implies \\
 \Delta_j &= j \left[ \frac{3 d}{4}  -\frac{3 \left( \sin \left(\frac{\pi  d}{4}\right)  \Gamma \left(\frac{d}{4}+1\right) \Gamma \left(\frac{3 d}{4}\right)\right)^{2/3}}{\pi ^{2/3} \Gamma \left(\frac{d}{2}\right)^{4/3}}\left( \frac{N + 8}{\epsilon j} \right)^{\frac{1}{3}} \right]
\end{split}
\end{equation}
 where we plugged in the fixed point value of $g$. At large $N$ these results agree with what we found before in \eqref{LargeNsmalljEps} and \eqref{LargeNlargejEps}.
  
\subsection{Large $N$ expansion}
We now revisit the large $N$ expansion for the long range $O(N)$ model from the cylinder approach. We start with the following action on $R \times S^{D - 1}$ 
\begin{equation}
S = \frac{\Gamma \left(\frac{s}{2} \right)}{(4 \pi)^{1 - \frac{s}{2}}} \int_{R \times S^{D - 1}}  \left( \frac{1}{2}  (\partial_{\mu} \Phi^I)^2 + \frac{(d - s)^2}{8}  (\Phi^I \Phi^I) \right) + \frac{1}{2} \int d^d x \sigma \phi^I \phi^I.
\end{equation} 
We take the same ansatz as in \eqref{ClassicalAnsatz} and the classical equation of motion and its solution are the same as in \eqref{ClassicalEOM} and \eqref{ClassicalEOMSoln}, but the boundary condition is now given by
\begin{equation}
\left[ \frac{ 4^{\frac{s}{2}} \Gamma \left(\frac{s}{2} \right)  (\sin \theta)^{d - 1} (\cos \theta)^{1 - s}}{\Gamma \left(1 - \frac{s}{2} \right)}  \partial_{\theta} f(\theta) +   c_{\sigma} f(\theta) \right]_{\theta = \frac{\pi}{2}} = 0.
\end{equation}
The $c_{\sigma}$ here is the classical value of $\sigma$, which is a constant and may be obtained by a Weyl transformation from \eqref{SigmaOnePFLat}. This boundary condition requires 
\begin{equation}
\frac{ \Gamma \left(\frac{d + s -2 \mu }{4} \right) \Gamma \left(\frac{d + s + 2 \mu }{4} \right) }{\Gamma \left(\frac{d - s - 2 \mu}{4}  \right) \Gamma \left(\frac{d - s + 2 \mu}{4} \right)} = - \frac{c_{\sigma}}{2^s}.
\end{equation}
Note that this is precisely the same as \eqref{EqWCSigma} that we found in the flat space approach. Using the boundary condition, one can see that the action actually vanishes on the classical solution. However, the effective action at large $N$ also involves the fluctuations. We expand the field around the classical background $\Phi^I = \Phi^I_{\textrm{cl}} + \delta \Phi^I $, and the the action up to quadratic order in fluctuations is given by 
\begin{equation}
S_{\textrm{fluct}} =  \frac{\Gamma \left(\frac{s}{2} \right)}{(4 \pi)^{1 - \frac{s}{2}}} \int_{R \times S^{D - 1}}  \left( \frac{1}{2}  (\partial_{\mu} \delta \Phi^I)^2 + \frac{(d - s)^2}{8}  (\delta \Phi^I  \delta \Phi^I) \right) + \frac{1}{2} \int d^d x \ c_{\sigma} \delta \phi^I \delta \phi^I.
\end{equation} 

To calculate the large $N$ free energy, we need to calculate the determinant of the fluctuations. One way to proceed is to reduce it down to $R \times S^{d - 1}$ again 
\begin{equation}
S = \frac{2 ^{s-1} \Gamma(\frac{d + s}{2})}{\pi^{\frac{d}{2}} \Gamma(-\frac{s}{2})} \int d^d x d^d y \sqrt{g_x} \sqrt{g_y} \frac{ \delta \phi^I (x) \delta \phi^I (y)}{(s(x,y))^{d+s}} + \frac{1}{2} \int d^d x \sqrt{g_x} \ c_{\sigma} \delta \phi^I \delta \phi^I
\end{equation} 
where $s(x,y)$ is the Weyl map of the flat space distance to the cylinder. Let us use the coordinates $(\tau, \vec{x})$ on the cylinder, then 
\begin{equation}
s(x, y)^2 = 2 \left( \cosh \left( \tau_x - \tau_y \right) -  \cos \theta \right)
\end{equation} 
where $\theta$ is the angle between $\vec{x}$ and $\vec{y}$ on the cylinder. The free energy on the cylinder is then given by  $\log \det \left( K \right) $ with $K$ defined by
\begin{equation}
K = \frac{2 ^{s} \Gamma(\frac{d + s}{2})}{\pi^{\frac{d}{2}} \Gamma(-\frac{s}{2})} \frac{1}{(s (x,y))^{d+s}} +  c_{\sigma} \frac{\delta^{d} (x - y)}{\sqrt{g_x}}.
\end{equation} 
Let us expand this operator into eigenfunctions of the Laplacian on the cylinder
\begin{equation}
\frac{1}{(s (x,y))^{d+s}} = \sum_{l, m} \int \frac{d \omega}{2 \pi} g (l, \omega) e^{i \omega (\tau_x - \tau_y)} Y^*_{l m} (\vec{x}) Y_{lm} (\vec{y})
\end{equation}
where $Y_{l m} $ are $d-1$ dimensional spherical harmonics and $e^{i \omega \tau}$ is the eigenfunction on the real line. Using orthogonality, we get 
\begin{equation} \label{EigenvalueIntegral}
g (l, \omega) = \frac{e^{ i \omega \tau_y}}{Y_{lm} (\vec{y})} \int d^d x \sqrt{g_x} \frac{1}{(s (x,y))^{d+s}} e^{- i \omega \tau_x} Y_{l m} (\vec{x})
\end{equation}
Note that the eigenvalue should not depend on $m$ because of the symmetries of $S^{d-1}$, so we can just evaluate it at $m = 0$. Using the same symmetries, we can fix $y$ to be at $\tau_y = 0$ and at the north pole of the sphere $S^{d - 1}$. Also note that $Y_{l0}(\theta)$ is proportional to the Gegenbauer polynomial $C_l^{(d-2)/2} (
\cos \theta)$. This results in the following integral 
\begin{equation} \label{EigenValueg}
\begin{split}
g (l, \omega) &= \frac{\textrm{Vol} (S^{d-2})}{C_l^{(d-2)/2} (1)}  \int d \tau d \theta \left( \sin \theta \right)^{d - 2} \frac{1}{( 2 \left( \cosh \tau  -  \cos \theta \right))^{\frac{d+s}{2}}} e^{- i \omega \tau} C_l^{(d-2)/2} (\cos \theta) 
\end{split}
\end{equation}
We can then use
\begin{equation}
\frac{1}{( 2 \left( \cosh \tau  -  \cos \theta \right))^{\frac{d+s}{2}}} = \sum_k C_{k}^{\frac{d + s}{2}} \left( \cos \theta \right) e^{- |\tau| \left( k + \frac{d + s}{2}  \right)}.
\end{equation} 
to turn the integral into a more useful form
\begin{equation}
g (l,\omega) = \frac{\textrm{Vol} (S^{d-2})}{C_l^{(d-2)/2} (1)} \sum_{k = 0}^{\infty}  \int_0^{\infty} d \tau 2 \cos \omega \tau e^{- \tau \left( k + \frac{d + s}{2}  \right)}  \int_{-1}^{1} d z \left(1  - z^2 \right)^{\frac{d - 3}{2}}  C_{k}^{\frac{d + s}{2}} \left( z \right) C_l^{(d-2)/2} (z).
\end{equation}
The integral over $\tau$ may be immediately performed. To perform the integral over $z$, we first use \eqref{GegenbauerIdentity} and then use the orthogonality relations to get
\begin{equation}
\begin{split}
g(l, \omega) = \frac{4 \pi^{d/2}}{\Gamma \left(\frac{d + s}{2} \right)} \sum_{k = 0}^{\infty} \frac{\Gamma \left(\frac{d + s}{2}+ l + k \right) \Gamma \left( 1 + \frac{s}{2}+ k \right) }{k! \Gamma \left(\frac{d}{2}+ l + k \right) \Gamma \left(1 + \frac{s}{2} \right) } \left( \frac{1}{(d + s + 2 l + 4k) + 2 i \omega} + \textrm{c.c.} \right)
\end{split}
\end{equation} 
The sum over $k$ can be computed in terms of generalized hypergeometric functions 
\begin{equation} \label{EigenValueRes}
\begin{split}
g(l, \omega) &= \frac{\pi ^{d/2} \Gamma (-s) \Gamma \left(\frac{d + s + 2 l}{2} \right)}{\Gamma \left(\frac{d + 2 l}{2}\right) \Gamma \left(\frac{d+s}{2}\right)} \bigg[ \frac{\Gamma \left(\frac{d + s + 2 l}{4} +\frac{i \omega}{2} \right)}{\Gamma \left(\frac{d - 3 s + 2 l}{4} +\frac{i \omega}{2}\right)}   \\ &\times \, _3F_2\left(\frac{d + s + 2 l}{4} +\frac{i \omega}{2} , \frac{d-s + 2 l}{2}-1,-\frac{s}{2};\frac{d + 2 l}{2},\frac{d - 3 s + 2 l}{4}+\frac{i \omega}{2} ;1\right)  + \textrm{c.c.} \bigg]. 
\end{split}
\end{equation}
For $s = 2$, this gives, as expected 
\begin{equation}
\frac{2 ^{s} \Gamma(\frac{d + s}{2})}{\pi^{\frac{d}{2}} \Gamma(-\frac{s}{2})} g (l, \omega) \bigg|_{s \rightarrow 2} =  \omega^2 + \left( \frac{d}{2}- 1 + l \right)^2. 
\end{equation}
The cylinder free energy may then be computed in terms of these eigenvalues 
\begin{equation}
\begin{split}
\mathcal{F} &= \frac{N}{2} \log \det \left( K \right) =  \frac{N}{2} \int d^d x \sqrt{g_x} \ \langle x | \log K  | x \rangle \\
&=  \frac{N}{2}  \sum_{l,m} \int \frac{d \omega}{2 \pi} \int d^d x \sqrt{g_x} \ |Y_{l,m} (\vec{x})|^2 \log \left( \frac{2 ^{s} \Gamma(\frac{d + s}{2})}{\pi^{\frac{d}{2}} \Gamma(-\frac{s}{2})} g (l, \omega)  +  c_{\sigma} \right). 
\end{split} 
\end{equation}
The spherical harmonics are normalized such that the integral over the sphere just gives one, while the integral over the real line gives the length of the cylinder. The sum over $m$ gives a factor of the degeneracy  
\begin{equation}
\begin{split}
\mathcal{F} &= \sum_{l } \frac{ N T (2 l + d - 2) \Gamma \left( l + d - 2 \right)}{ 2 l! \Gamma \left(d - 1 \right)} \int \frac{d \omega}{2 \pi}  \log \left( \frac{2 ^{s} \Gamma(\frac{d + s}{2})}{\pi^{\frac{d}{2}} \Gamma(-\frac{s}{2})} g (l, \omega)  +  c_{\sigma} \right) \\
&= \sum_{l } \frac{ N T (2 l + d - 2) \Gamma \left( l + d - 2 \right)}{ 2 l! \Gamma \left(d - 1 \right)} \int \frac{d \omega}{ 2 \pi} \bigg[  \log \left( \frac{2^{s} \Gamma(\frac{d + s}{2})}{\pi^{\frac{d}{2}} \Gamma(-\frac{s}{2})} g (l, \omega) \right)  \\
& \hspace{7cm} + \log \left( 1 + \frac{\pi^{\frac{d}{2}} \Gamma(-\frac{s}{2}) }{ 2^{s} \Gamma(\frac{d + s}{2}) g (l, \omega)}  c_{\sigma} \right) \bigg] .
\end{split}
\end{equation}
In the second line we separated out the $c_{\sigma} = 0$ piece of the free energy. This is the vacuum energy, which should be subtracted while computing the scaling dimensions. After this subtraction, the scaling dimensions may be calculated as in the previous subsection, by extremizing the following expression with respect to $\mu$ 
\begin{equation}
\begin{split}
\Delta_j &= \frac{\mathcal{F}}{T} + \mu j\\
&= N \left[ \sum_{l} \frac{ (2 l + d - 2) \Gamma \left( l + d - 2 \right)}{l! \Gamma \left(d - 1 \right)} \int \frac{d \omega}{ 4 \pi} \log \left( 1 + \frac{\pi^{\frac{d}{2}} \Gamma(-\frac{s}{2}) }{ 2^{s} \Gamma(\frac{d + s}{2}) g (l, \omega)}  c_{\sigma} \right) + \mu \hat{j}  \right]
\end{split}
\end{equation}
For equivalence to the flat space calculation in \eqref{ScalingDimRes}, we want the first term to be identified with $F(c_{\sigma})$ in \eqref{FunctDetRes} which requires 
\begin{equation}
g (l, \omega) = \frac{\Gamma \left(-\frac{s}{2} \right) \pi^{\frac{d}{2}} \Gamma \left(\frac{s}{2} \right) }{ \Gamma \left( \frac{d + s}{2} \right) \Gamma \left( \frac{d - s}{2} \right)  Q_{l} \left(\frac{\omega}{2} \right) } .
\end{equation}
Comparing \eqref{IntegralResult} and \eqref{EigenValueRes},  we need the following hypergeometric identity to hold 
\begin{equation}
\begin{split}
&\frac{\, _3 F_2(a,b,c;a-b+1,2 b+c;1)}{\Gamma (a-2 b-c+1) \Gamma (a- b + 1) \Gamma (c + 2 b) } + \frac{\, _3 F_2(a,b,a-2 b-c+1;a-b+1,a-c+1;1)}{\Gamma (c) \Gamma (a- b + 1) \Gamma (a- c + 1)} \\
&= \frac{\sqrt{\pi } 2^{1-2 b}}{\Gamma \left(b+\frac{1}{2}\right) \Gamma (a-2 b+1) \Gamma (b+c) \Gamma (a-b-c+1)}
\end{split}
\end{equation}
We could not prove this identity or find it in the literature, but we checked that it holds numerically for a wide range of parameters, so we expect it to be true. Hence, as promised, we have shown that the scaling dimensions calculated on $R^d$ and from the cylinder approach match. 

\section*{Acknowledgments}
This research was supported in part by the US NSF under Grant No.~PHY-1914860.

\appendix

\section{Scaling dimensions from standard $1/N$ perturbation theory}  \label{App:Pert1N}
\begin{figure} 
\centering
\begin{subfigure}{0.15\textwidth}
\includegraphics[width = \textwidth]{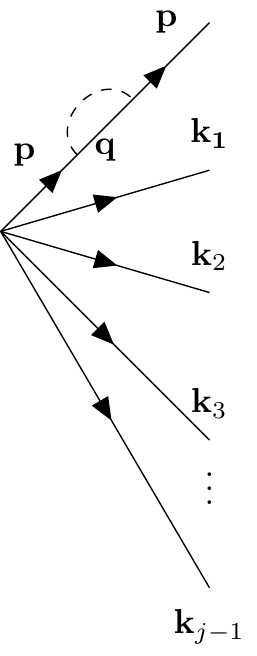}
\end{subfigure} \hspace{1cm}
\begin{subfigure}{0.15\textwidth}
\includegraphics[width = \textwidth]{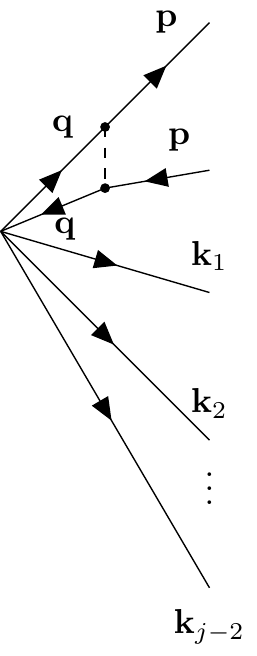}
\end{subfigure}
\caption{The two diagrams that contribute to the $1/N$ correction to the correlator. The dashed line represents the $\sigma$ propagator. }
\label{FigureAnomFeynman}
\end{figure}

In the regime when $j \ll N$ or $\hat{j} \ll 1$, we can use ordinary $1/N$ perturbation theory to calculate the scaling dimensions of the operators $\co_j$. We will do that in this appendix, and it will serve as a check of the calculations in section \ref{Sec:ONFlat}. We will calculate the correlator of the operator $\co_j$ with $j$ fundamental fields $\phi$. Let us look at this correlator in momentum space. Just by dimensional analysis \footnote{We are suppressing $O(N)$ indices in this appendix} 
\begin{equation}
\langle \co_j(0) \phi(\mathbf{k}_1) \dots \phi(\mathbf{k}_1) \rangle = \tilde{G}(k_1,\dots,k_{j}) \propto \dfrac{1}{|p|^{jd - \Delta_j - j\Delta_\phi}}. 
\end{equation}
The momentum conservation requires $\sum_j \mathbf{k}_j = 0$. The last term is just schematic and is meant to count the powers of momentum. It is well known that the field $\phi$ in the long range model does not receive anomalous dimensions \cite{PhysRevLett.29.917, Giombi:2019enr}
\begin{equation}
\Delta_\phi = \dfrac{d-s}{2}, \hspace{1cm} \Delta_j = j\lr{\dfrac{d-s}{2}} + \dfrac{\gamma_j}{N} + O \left(\frac{1}{N^2} \right). 
\end{equation}
Therefore all the logarithmic terms in the correlator must contribute to the anomalous dimensions of $\co_j$ 
\beq
\tilde{G}(k_1,\dots,k_{j}) \propto \dfrac{1}{|p|^{js - \gamma_j /N}} = \dfrac{1}{|p|^{js}} \lr{1+ \dfrac{\gamma_j}{2N} \log(p^2) + \co(\tfrac{1}{N^2})}.
\eeq
There are 2 types of diagrams that contribute to the $1/N$ correction to this $(j+1)$-point function as shown in figure \ref{FigureAnomFeynman}.  

The left diagram does not contribute to the anomalous dimension because it does not give rise to any $\log(p^2)$ terms. This is also the reason why $\phi$ does not get anomalous dimensions. 
\begin{equation}
\begin{split}
= \dfrac{1}{|k_1|^s \cdots |k_{j-2}|^s} &\dfrac{\delta_{IJ} \tilde{C}_\sigma}{N |p|^{2s}} \int \dfrac{d^d q}{(2\pi)^d} \dfrac{1}{(q^2)^{s/2} ((p-q)^2)^{d/2-s}} \\
= \dfrac{1}{|k_1|^s \cdots |k_{j-2}|^s}&\dfrac{\delta_{IJ} \tilde{C}_\sigma}{(4\pi)^{d/2} N |p|^s}\dfrac{\Gamma(-\tfrac{s}{2}) \Gamma(\tfrac{d-s}{2}) \Gamma(s)}{\Gamma(\halfd -s)\Gamma(\tfrac{s}{2}) \Gamma(\tfrac{d+s}{2})}
\end{split}
\end{equation}
We used that the $\sigma$ propagator is given by 
\begin{equation}
\langle \sigma (-\mathbf{q}) \sigma (\mathbf{q}) \rangle = \frac{\tilde{C}_\sigma}{N(q^2 )^{\frac{d}{2} - s}}, \hspace{1cm} \tilde{C_{\sigma}} = -\frac{2 (4 \pi)^{\frac{d}{2}} \Gamma(\frac{s}{2})^2 \Gamma(d-s) }{\Gamma (s - \frac{d}{2}) \Gamma(\frac{d-s}{2})^2} .
\end{equation}

On the other hand, the right diagram does give rise to logarithms. There are ${j \choose 2} = j(j-1)/2$ diagrams of this form. For simplicity, we picked a configuration such that $\mathbf{k}_1 = - \mathbf{k}_2 = \mathbf{p}$ where the diagram may be computed by the following simple integral
\begin{equation} \label{eq:jplus1}
 \dfrac{1}{|k_1|^s \cdots |k_{j-2}|^s} \delta_{IJ} \int \dfrac{d^d q}{(2\pi)^d} \dfrac{1}{|p|^{2s}|q|^{2s}} \dfrac{\tilde{C}_\sigma}{N}\dfrac{1}{((q-p)^2)^{d/2-s}} 
\end{equation}
The integral can be computed using Feynman parameters and introducing a regulator $\eta$. 
\begin{equation}
\begin{split}
\int \dfrac{d^d q}{(2\pi)^d} &\dfrac{1}{|p|^{2s}|q|^{2s}} \dfrac{\tilde{C}_\sigma}{N}\dfrac{1}{((q-p)^2)^{d/2-s}}  \\
&=\dfrac{  \tilde{C}_\sigma}{N |p|^{2s}} \dfrac{\Gamma(\halfd)}{\Gamma(s)\Gamma(\halfd-s)}  \int d\alpha \alpha^{s-1}(1-\alpha)^{d/2-s-1} \int \dfrac{d^d q}{(2\pi)^d} \dfrac{1}{[q^2 +\alpha(1-\alpha)p^2]^{d/2}} \\
&=  \dfrac{ \tilde{C}_\sigma}{N |p|^{2s}} \dfrac{\Gamma(\halfd)}{\Gamma(s)\Gamma(\halfd-s)}  \int d\alpha \alpha^{s-1}(1-\alpha)^{d/2-s-1}  \dfrac{2\pi^{d/2}}{\Gamma(\halfd)}\int \dfrac{d q}{(2\pi)^d} \dfrac{q^{d-1-\eta}}{[q^2 +\alpha(1-\alpha)p^2]^{d/2}} \\
&=  \dfrac{  \tilde{C}_\sigma (4\pi)^{-d/2}}{N |p|^{2s} \Gamma(s) \Gamma(\halfd-s)\Gamma(\halfd)}\left[ \co(\tfrac{1}{\eta}) - \Gamma(\halfd-s) \Gamma(s) (\log(p^2) + \dots) + \co(\eta)\right] 
\end{split}
\end{equation}
Dropping the $1/\eta$ pole and taking the limit $\eta \to 0$, we see that the coefficient of the $\log (p^2)$ piece, after summing all the diagrams of the form \ref{eq:jplus1} is  
\beq
\dfrac{1}{|k_1|^s \cdots |k_{j-2}|^s |p|^{2s}} \dfrac{ -\tilde{C}_\sigma j(j-1)}{(4\pi)^{d/2} \Gamma(\halfd)}\dfrac{1}{2N}
\eeq
Thus the dimension of $\co_j$ is given by
\beq
\Delta_j =  \lr{\frac{d-s}{2}} j + \dfrac{2 \Gamma(d-s) \Gamma(\frac{s}{2})^2 j (j - 1)}{\Gamma(\frac{d-s}{2})^2 \Gamma(s-\halfd)\Gamma(\halfd) N }  + O \left(\frac{1}{N^2} \right). 
\eeq
At large $j$ this is consistent with what we found in \eqref{SmalljSaddSol}.

\section{Large $c_{\sigma}$ expansion of $F(c_{\sigma})$ using heat kernel methods} \label{App:largecF}
In this appendix, we show that at large $c_{\sigma}$, the functional determinant $F(c_{\sigma})$ behaves as \eqref{LargecF}. We will use heat kernel methods to calculate the functional determinant (see for instance \cite{Schubert:2001he, Vassilevich:2003xt, Bastianelli:2006hq} for reviews). We start with the following representation of the functional determinant
\begin{equation}
\frac{1}{2} \textrm{Tr} \log \left( (-\nabla^2)^{\frac{s}{2}} + \sigma_* (x) \right) = - \frac{1}{2} \int d^d x \langle x | \int_0^{\infty} \frac{d T}{T} e^{- T \left( P^s + \sigma^*(X) \right)} | x \rangle 
\end{equation}
where we use capital letters to denote operators. Recall that 
\begin{equation}
\sigma^*(x) = c_{\sigma} \frac{|x_1 - x_2|^s}{|x_1 - x|^s |x_2 - x|^s} \equiv c_{\sigma} V(x).
\end{equation}
We then use Trotter formula to write the determinant as a path integral. For that, we divide $T$ into $N$ pieces, then for $N$ very large, we may write 
\begin{equation} 
\begin{split}
&\frac{1}{2} \textrm{Tr} \log \left( (-\nabla^2)^{\frac{s}{2}} + \sigma_* (x) \right) = - \frac{1}{2} \int d^d x_0 \int_0^{\infty} \frac{d T}{T} \langle x_0 | \left( e^{- \frac{T}{N}  P^s} e^{ - \frac{T}{N} \sigma^*(X) } \right)^N | x_0 \rangle \\
&=  - \frac{1}{2} \int_0^{\infty} \frac{d T}{T} \int_{x(0) = x(T) = x_0} D x(\tau) D p(\tau) e^{\int_0^T d \tau \left( i p(\tau) \cdot \dot{x}(\tau) - (p^2)^{s/2}(\tau) - c_{\sigma} V(x)  \right)} \\
&= - \frac{1}{2} \int_0^{\infty} \frac{d T}{T} \int_{x(0) = x(1) = x_0} D x(t) D p(t) e^{\int_0^1 d t \left( i p(t) \cdot \dot{x}(t) - T (p^2)^{s/2}(t) -  T c_{\sigma} V(x)  \right)}.
\end{split}
\end{equation}

To get the large $c_{\sigma}$ behavior, we rescale $T \rightarrow T / c_{\sigma}$ and at the same time, also rescale $p \rightarrow c_{\sigma}^{1/s} p $ to get 
\begin{equation}
\frac{1}{2} \textrm{Tr} \log \left( (-\nabla^2)^{\frac{s}{2}} + \sigma_* (x) \right) = - \frac{1}{2} \int_0^{\infty} \frac{d T}{T} \int D x(t) D [ c_{\sigma}^{1/s} p(t) ] e^{\int_0^1 d t \left( i c_{\sigma}^{1/s} p(t) \cdot \dot{x}(t) - T (p^2)^{s/2}  -  T V(x)  \right)}.
\end{equation}
At large $c_{\sigma}$, the path integral will be dominated by constant $x$ configurations. A path integral over $x$ fluctuations will then also force momenta to be constant at large $c_{\sigma}$. So we can expand about the constant $x$ and $p$ configurations to obtain an expansion in $1/c_{\sigma}$
\begin{equation}
\begin{split}
x(t) &= x_0 + \frac{1}{c_{\sigma}^{1/s} } \chi(t), \hspace{1cm} p(t) = p_0 + \Pi(t) \implies \\
 \frac{1}{2} \textrm{Tr} \log \left(...\right) &= - \frac{1}{2} \int_0^{\infty} \frac{d T}{T} \int \frac{d^d x_0 d^d p_0 c_{\sigma}^{d/s} }{(2 \pi)^d} e^{- T (p_0^2)^{s/2} } e^{- T V(x_0)} \int D \chi(t) D \Pi(t) e^{-S} 
\end{split}
\end{equation}
where the action to quadratic order in fluctuations is given by 
\begin{equation}
\begin{split}
S = \int_0^1 d t \bigg[ & - i \Pi(t) \cdot \dot{\chi}(t) - \frac{T p_0^{s - 2} s }{2} \left( \Pi^2 + 2 p_0 \cdot \Pi \right) +  \frac{T p_0^{s - 4} s (s - 2) }{2} \left( p_0 \cdot \Pi \right)^2 \\
&  + \frac{T}{c_{\sigma}^{1/s}} \chi^{\mu} \partial_{\mu} V + \frac{T}{2 c_{\sigma}^{2/s}} \chi^{\mu} \chi^{\nu} \partial_{\mu} \partial_{\nu} V  \bigg].
\end{split}
\end{equation}
First, note that at leading order in $c_{\sigma}$ we can ignore the fluctuations, and then we can do the integral over $p_0$ followed by an integral over $T$ and finally over $x_0$
\begin{equation}
\begin{split}
\frac{1}{2} \textrm{Tr} \log \left( ... \right) &=  - \frac{1}{2} \int_0^{\infty} \frac{d T}{T} \int \frac{d^d x_0 d^d p_0 c_{\sigma}^{d/s} }{(2 \pi)^d} e^{- T (p_0^2)^{s/2} } e^{- T V(x_0)} \\
&= - \frac{\Gamma \left( \frac{d}{s} \right) \Gamma \left(- \frac{d}{s} \right) c_{\sigma}^{d/s} }{s (4 \pi)^{d/2} \Gamma \left( \frac{d}{2} \right) } \int d^d x_0 \frac{|x_1 - x_2|^d}{|x_0 - x_1|^d |x_0 - x_2|^d} \\
& = - \frac{(c_{\sigma})^{\frac{d}{s}} \pi  }{2^{d - 1} \ d  \ \Gamma \left( \frac{d}{2} \right)^2 \sin \left( \frac{\pi d}{s} \right)}\log \left( \frac{\delta^2}{ |x_{12}|^2} \right)
\end{split}
\end{equation}
Note that the last integral over $x_0$ is the same as in \eqref{DivIntegral}. To calculate corrections to it, we expand the fluctuations into Fourier modes 
\begin{equation}
\chi^{\mu}(t) = \sum_{m= 1}^{\infty} \left(  \chi^{\mu}_m  \sin (2\pi m t) + \tilde{\chi}^{\mu}_m  \cos (2\pi m t)  \right), \hspace{0.2cm} \Pi^{\mu}(t) = \sum_{m= 1}^{\infty}   \Pi^{\mu}_m  \sin (2\pi m t) + \tilde{\Pi}^{\mu}_m  \cos (2\pi m t)  .
\end{equation}
The action in terms of these modes is given by 
\begin{equation} \label{ActionModeFluct}
\begin{split}
S =  \sum_{m} \bigg[& -i \pi m \left( \tilde{\Pi}_m \cdot \chi_m -  \Pi_m \cdot \tilde{\chi}_m \right) + \frac{T}{4 c_{\sigma}^{2/s}} \left( \chi_m^{\mu} \chi_m^{\nu} +  \tilde{\chi}_m^{\mu} \tilde{\chi}_m^{\nu}  \right) \partial_{\mu} \partial_{\nu} V  + \\
&+ T \left(\frac{s p_0^{s - 2}}{4} \left(\Pi_m \cdot \Pi_m +  \tilde{\Pi}_m \cdot \tilde{\Pi}_m \right) + \frac{s (s - 2) p_0^{s - 4} }{4}  \left( (p_0 \cdot \Pi_m)^2 + (p_0 \cdot \tilde{\Pi}_m)^2 \right) \right) \bigg].
\end{split}
\end{equation}
The last term in the above action mixes the $\Pi$ modes in different directions, and is therefore slightly tedious to deal with. 
Let us start with the case when $d = 1$, so there is only one direction and no mixing. Then the action simplifies to 
\begin{equation}
\begin{split}
S =  \sum_{m} \bigg[& -i \pi m \left( \tilde{\Pi}_m  \chi_m -  \Pi_m \tilde{\chi}_m \right) + \frac{T}{4 c_{\sigma}^{2/s}} \left( \chi_m^{2} +  \tilde{\chi}_m^{2}  \right) \partial^2 V  +  \frac{ T s (s - 1) p_0^{s - 2}}{4} \left(\Pi_m^2 +  \tilde{\Pi}_m^2 \right) \bigg].
\end{split}
\end{equation}
Then the path integral over $\Pi$ and $\chi$ may be easily performed 
\begin{equation} 
\begin{split}
\int D \chi(t) D \Pi(t) e^{-S}  &= \prod_m \int d \chi_m d \tilde{\chi}_m \left( \frac{\pi m^2}{T s (s - 1) p_0^{s - 2} } \right) e^{ -  \frac{\pi^2 m^2}{T s (s - 1) p_0^{s - 2}} \left( \chi_m^{2} +  \tilde{\chi}_m^{2}  \right) -  \frac{T}{4 c_{\sigma}^{2/s}} \left( \chi_m^{2} +  \tilde{\chi}_m^{2}  \right) \partial^2 V  } \\
& = 1 - \frac{T^2 \partial^2 V s (s - 1) p_0^{s - 2}  }{24 c_{\sigma}^{2/s}}.
\end{split}
\end{equation}
We chose the path integral measure such that the path integral is normalized to one when the potential vanishes. The functional determinant is then given by 
\begin{equation}
\begin{split}
\frac{1}{2} \textrm{Tr} \log \left( ... \right) &=  - \frac{1}{2} \int_0^{\infty} \frac{d T}{T} \int \frac{d x_0 d p_0 c_{\sigma}^{1/s} }{2 \pi} e^{- T (p_0^2)^{s/2} } e^{- T V(x_0)} \left( 1 - \frac{T^2 \partial^2 V s (s - 1) p_0^{s - 2}  }{24 c_{\sigma}^{2/s}} \right) \\
&= - \int_0^{\infty} \frac{d T}{T^{\frac{1}{s} + 1}} \int \frac{d x_0 c_{\sigma}^{1/s} \Gamma \left( \frac{1}{s} \right) }{2 \pi s} e^{- T V(x_0)} \left( 1 - \frac{T^{\frac{2}{s} + 1} \partial^2 V s (s - 1)  \Gamma \left(1 -  \frac{1}{s} \right) }{24 c_{\sigma}^{2/s} \Gamma \left( \frac{1}{s} \right)} \right) \\
&= - \int \frac{d x_0 c_{\sigma}^{1/s} V^{1/s} \Gamma \left( \frac{1}{s} \right) \Gamma \left(- \frac{1}{s} \right) }{2 \pi s}   \left( 1 - \frac{ (1 - s) \partial^2 V }{24 \ s \  c_{\sigma}^{2/s} V^{\frac{2}{s} + 1} }  \right).
\end{split}
\end{equation}
Recall that 
\begin{equation}
V(x_0) = \frac{|x_1 - x_2|^s}{|x_0 - x_1|^s |x_0 - x_2|^s} \implies \frac{\partial^2 V }{V^{\frac{2}{s} + 1}} = s  (s + 2 - d) + 2 s (2 s + 2 - d) \frac{(x_0 - x_1) \cdot (x_0 - x_2)}{|x_1 - x_2|^2}.
\end{equation}
But the second term above, when multiplied by $V^{d/s}$, is proportional to a total derivative. This can be seen from the following
\begin{equation}
\frac{ (x_0 - x_1) \cdot (x_0 - x_2)}{|x_0 - x_1|^d |x_0 - x_2|^d} = -\frac{1}{2 (d - 2)} \partial_i \left( \frac{(x_0 - x_1)^i}{|x_0 - x_1|^d |x_0 - x_2|^{d - 2}} + \frac{(x_0 - x_2)^i}{|x_0 - x_1|^{d - 2} |x_0 - x_2|^{d}} \right)
\end{equation}
so it does not contribute to the integral. Therefore $\partial^2 V$ term only changes the $x_0$ integral by a constant factor. The integral over $x_0$ may then be easily performed by using \eqref{DivIntegral} 
\begin{equation}
\begin{split}
\frac{1}{2} \textrm{Tr} \log \left( ... \right) &=  - \int \frac{d x_0 c_{\sigma}^{1/s} V^{1/s} \Gamma \left( \frac{1}{s} \right) \Gamma \left(- \frac{1}{s} \right) }{2 \pi s}   \left( 1 - \frac{(1 - s^2)}{24  c_{\sigma}^{2/s} }  \right) \\ &= - \frac{c_{\sigma}^{1/s}}{\sin \left( \frac{\pi}{s} \right)} \left( 1 - \frac{(1 - s^2)}{24  c_{\sigma}^{2/s} }  \right) \log \left( \frac{\delta^2}{ |x_{12}|^2} \right).
\end{split}
\end{equation}
This implies that in $d = 1$, we get 
\begin{equation} \label{FcLargec1dApp}
F(c_{\sigma}) = \frac{c_{\sigma}^{1/s}}{\sin \left( \frac{\pi}{s} \right)} \left( 1 - \frac{(1 - s^2)}{24  c_{\sigma}^{2/s} }  \right)  
\end{equation}

Another case when there is no mixing in \eqref{ActionModeFluct} is $s = 2$ for any $d$ when the last term in \eqref{ActionModeFluct} vanishes. In that case also, the path integral over $\chi$ and $\Pi$ may be done 
\begin{equation} 
\begin{split}
\int D \chi(t) D \Pi(t) e^{-S}  &= \prod_{m, \mu} \int d \chi_m d \tilde{\chi}_m \left( \frac{\pi m^2}{2 T} \right) e^{ -  \frac{\pi^2 m^2}{2 T } \left( \chi_m^{2} +  \tilde{\chi}_m^{2}  \right) -  \frac{T}{4 c_{\sigma}} \left( \chi_m^{2} +  \tilde{\chi}_m^{2}  \right) \partial^2 V  } \\
& = 1 - \frac{T^2 \partial^2 V }{12 c_{\sigma}}.
\end{split}
\end{equation}
We can then calculate the functional determinant by integrating over $p_0, T$ and $x_0$ as before 
\begin{equation}
\begin{split}
 \frac{1}{2} \textrm{Tr} \log \left(...\right) &= - \frac{1}{2} \int_0^{\infty} \frac{d T}{T} \int \frac{d^d x_0 d^d p_0 c_{\sigma}^{d/2} }{(2 \pi)^d} e^{- T p_0^2 } e^{- T V(x_0)} \left(1 - \frac{T^2 \partial^2 V }{12 c_{\sigma}} \right) \\
&= - \frac{ \Gamma\left(- \frac{d}{2} \right) (c_{\sigma})^{d/2}}{2} \int \frac{d^d x_0 V^{d/2} }{(4 \pi)^{d/2}} \left[ 1 + \frac{d (d - 2) (d - 4) }{24 \ c_{\sigma}}  + O \left(\frac{1}{c_{\sigma}^2} \right)  \right] \\
& =  \frac{\Gamma\left(- \frac{d}{2} \right) (c_{\sigma})^{d/2} }{2^d \Gamma\left(\frac{d}{2} \right)} \left[ 1 + \frac{d (d - 2) (d - 4) }{24 \ c_{\sigma}}  + O \left(\frac{1}{c_{\sigma}^2} \right)  \right] \log \left( \frac{\delta^2}{ |x_{12}|^2} \right)
\end{split}
\end{equation}
which then implies 
\begin{equation}
F(c_{\sigma}) =  - \frac{\Gamma\left(- \frac{d}{2} \right) (c_{\sigma})^{d/2} }{2^d \Gamma\left(\frac{d}{2} \right)} \left[ 1 + \frac{d (d - 2) (d - 4) }{24 \ c_{\sigma}}  + O \left(\frac{1}{c_{\sigma}^2} \right)  \right].   
\end{equation}
This agrees with what was found in \cite{Giombi:2020enj}. We will not do the general calculation for general $d$ and $s$, but from the  structure of \eqref{ActionModeFluct}, we expect the corrections to the leading large $c_{\sigma}$ behavior to be of order $1/c_{\sigma}^{2/s}$, so that 
\begin{equation} \label{FcsLargecApp}
F(c_{\sigma}) = \frac{(c_{\sigma})^{\frac{d}{s}} \pi  }{2^{d - 1} \ d  \ \Gamma \left( \frac{d}{2} \right)^2 \sin \left( \frac{\pi d}{s} \right)} \left( 1 + O \left( \frac{1}{c_{\sigma}^{2/s}} \right) \right). 
\end{equation}

\bibliographystyle{ssg}
\bibliography{LongRange-bib}

\end{document}